\documentclass[pre,amsmath,amssymb,superscriptaddress,graphicx,showkeys,showpacs,reprint,twocolumn]{revtex4-2}

\usepackage[colorlinks=true,linkcolor=blue,citecolor=blue,bookmarks]{hyperref}
\usepackage{url}
\usepackage{units}
\usepackage{graphicx}
\usepackage{subfigure}
\usepackage[utf8]{inputenc}
\usepackage{ae}
\usepackage{natbib}
\usepackage{dcolumn}
\usepackage{multirow}
\usepackage{dsfont}
\usepackage{amssymb,amsmath,amsfonts,amsthm}
\usepackage{comment}

\def\R{\mathbb{R}}
\def\N{\mathbb{N}}

\newcommand{\E}[1]{\mathbb{E}\left[{#1}\right]}

\newcommand{\U}{\mathcal{U}_{PR}^{c}}
\newcommand{\1}{\mathds{1}}
\newcommand{\0}{\mathbf{0}}

\newcommand{\SKP}[2]{\langle #1 , #2 \rangle}
\renewcommand{\d}[1]{ \mathrm{d}{#1} }

\DeclareMathOperator{\reach}{reach}

\newtheorem{Th}{Theorem}[section]
\newtheorem{Prop}[Th]{Proposition}
\newtheorem{Lem}[Th]{Lemma}

\newtheorem{Alg}[Th]{Algorithm}

\newcolumntype{d}{D{.}{.}{7}}

\begin{document}



\title{Minkowski Correlation Functions (MCF):\\ Long-range, robust shape descriptors for point patterns}

\author{Dominik Pabst}
\affiliation{Department of Physics, Friedrich-Alexander-Universität Erlangen-Nürnberg (FAU), Staudtstraße 7, D-91058 Erlangen, Germany}
\author{Michael A. Klatt}
\affiliation{German Aerospace Center (DLR), Institute for AI Safety and Security, Wilhelm-Runge-Str. 10, 89081 Ulm, Germany; German Aerospace Center (DLR), Institute for Material Physics in Space, 51170 Köln, Germany; Department of Physics, Ludwig-Maximilians-Universität München, Schellingstr. 4, 80799 Munich, Germany}
\author{Klaus Mecke}
\affiliation{Department of Physics, Friedrich-Alexander-Universität Erlangen-Nürnberg (FAU), Staudtstraße 7, D-91058 Erlangen, Germany}

\date{\today}

\begin{abstract}
A morphometric analysis of point patterns using Minkowski functionals is short-sighted, whereas correlation functions do not provide robust, i.e., additive shape descriptors. To combine the advantages of both standard techniques and to overcome their weaknesses, we introduce ring-based Minkowski correlation functions (MCFs) as robust and sensitive descriptors of disordered point patterns.
By assigning to each point a spherical shell and quantifying the resulting two-phase morphology via Minkowski functionals, MCFs capture structural information across multiple length scales, remain sensitive at radii where conventional disk-based Minkowski functionals have saturated, and implicitly encode higher-order correlations beyond the pair correlation and structure factor.
We present an exact and efficient algorithm for planar patterns extending the established disk-based morphological approach to long ranges.
Applications across scales -- active colloids (Quincke rollers), longleaf pine trees, and the Hubble Ultra Deep Field -- demonstrate clear and systematic deviations from homogeneous Poisson references, with distinct responses of covered area, boundary length, and Euler characteristic.
The approach combines short computation times, robustness to noise, and straightforward edge handling with clear interpretability as low-dimensional functions of the outer ring radius.
\end{abstract}

\pacs{}{}
\maketitle

\section{Introduction}
\label{Introduction}

From astronomical sources in the Hubble Ultra Deep Field to longleaf pine trees and microscopic Quincke rollers, disordered point patterns appear ubiquitously in nature on nearly all length scales.
Efficient and sensitive structure characteristics are needed for physical or biological insights, e.\,g., to explain the mechanism of their formation, as well as for statistical data analysis or model selection.
Moreover, these measures should be robust against noise and even with limited statistics they should be able to detect correlations in the point configurations like clustering or local crystallization.

Common tools are the pair correlation function $g^{(2)}(\mathbf{x}_{1},\mathbf{x}_2)$ or more generally the so-called standard $n$-point correlation functions $g^{(n)}(\mathbf{x}_1,\mathbf{x}_2,\ldots \mathbf{x}_n)$.
They describe the joint probability densities of finding points in infinitesimal neighborhoods of the positions $\mathbf{x}_1,\mathbf{x}_2,\ldots,\mathbf{x}_n$, normalized by the $n$-th power of the intensity.
They are, for example, used in materials science~\cite{Torquato1982, Torquato2002}, in medical image processing~\cite{MosaligantiEtAl2008}, and astronomy~\cite{White1979, Peebles1980, SzapudiEtAl2001, CameronEtAl2011}.
Recent extensions of correlation-based summary statistics include local mark correlation functions that resolve spatially varying mark associations in marked point processes~\cite{Eckardt2026}.
However, the standard pair correlation function is by definition limited to two-point information, and multipoint correlation functions quickly become intractable both analytically and numerically.
The brute-force approach to calculate an $n$-point correlation function scales with the number of data points $N$ like $\mathcal{O}(N^n)$.
This can be critical for even modest-sized data sets.
Although complex algorithms have been developed to improve upon this scaling, they typically still allow only for estimates of the three- or four-point correlation functions~\cite{MooreEtAl2001, ZhangPen2005, ZhangYu2011, MarchEtAl:2012}.

\begin{figure}[t]
  \centering
  \begin{minipage}[c][][c]{0.32\linewidth}
    {\includegraphics[width=\linewidth]{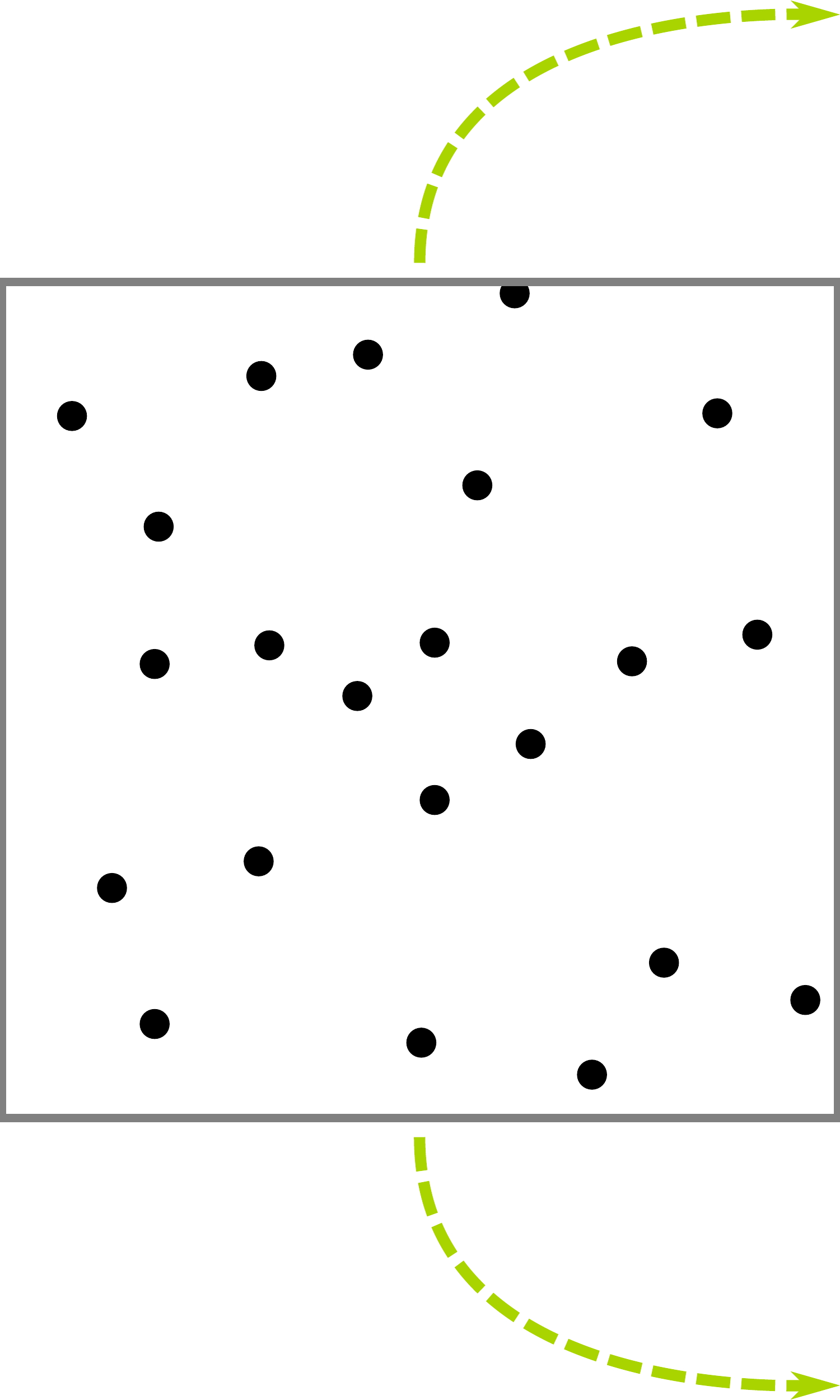}}
  \end{minipage}
  \begin{minipage}[c][][c]{0.64\linewidth}
    \subfigure[][]{\includegraphics[width=0.49\linewidth]{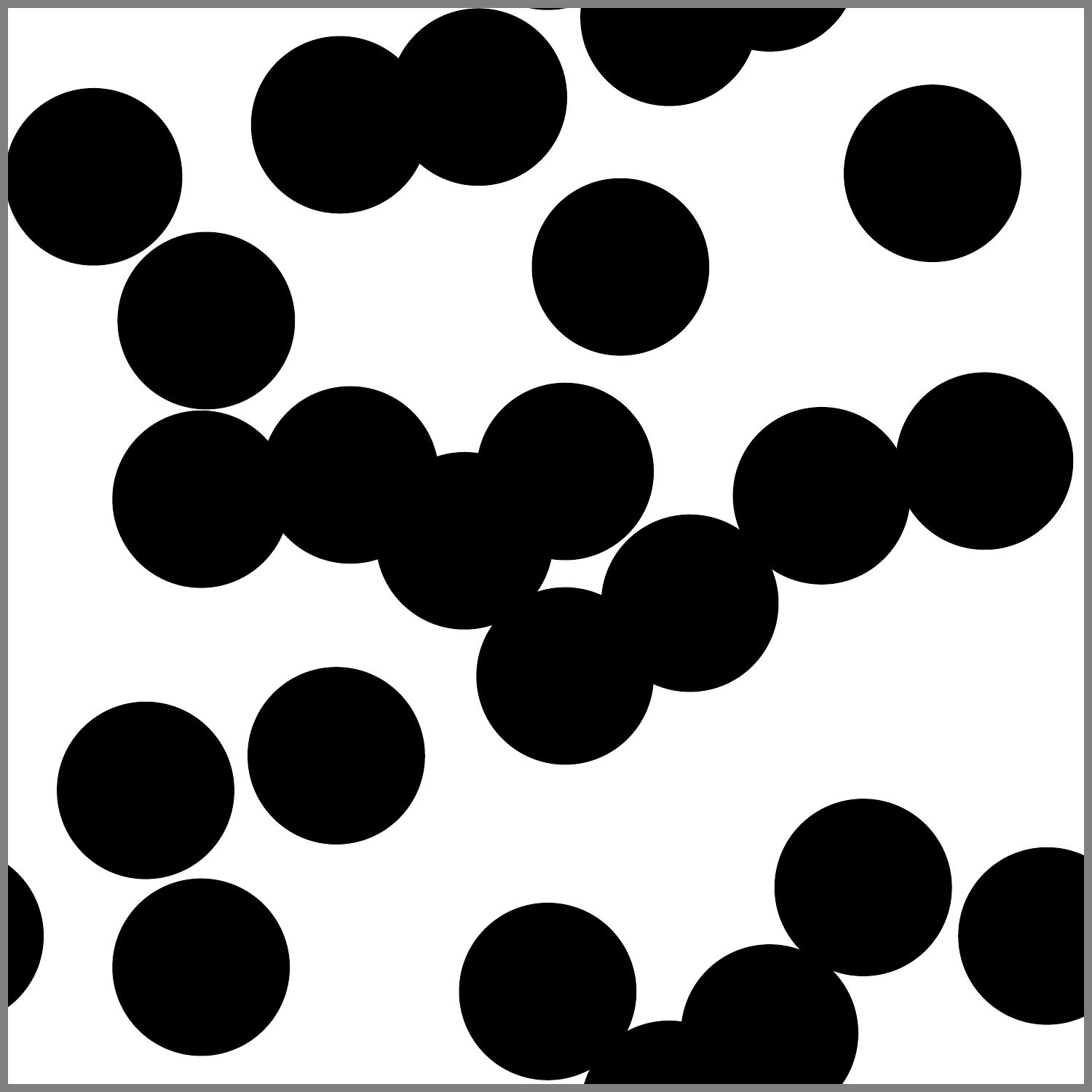}}
    \subfigure[][]{\includegraphics[width=0.49\linewidth]{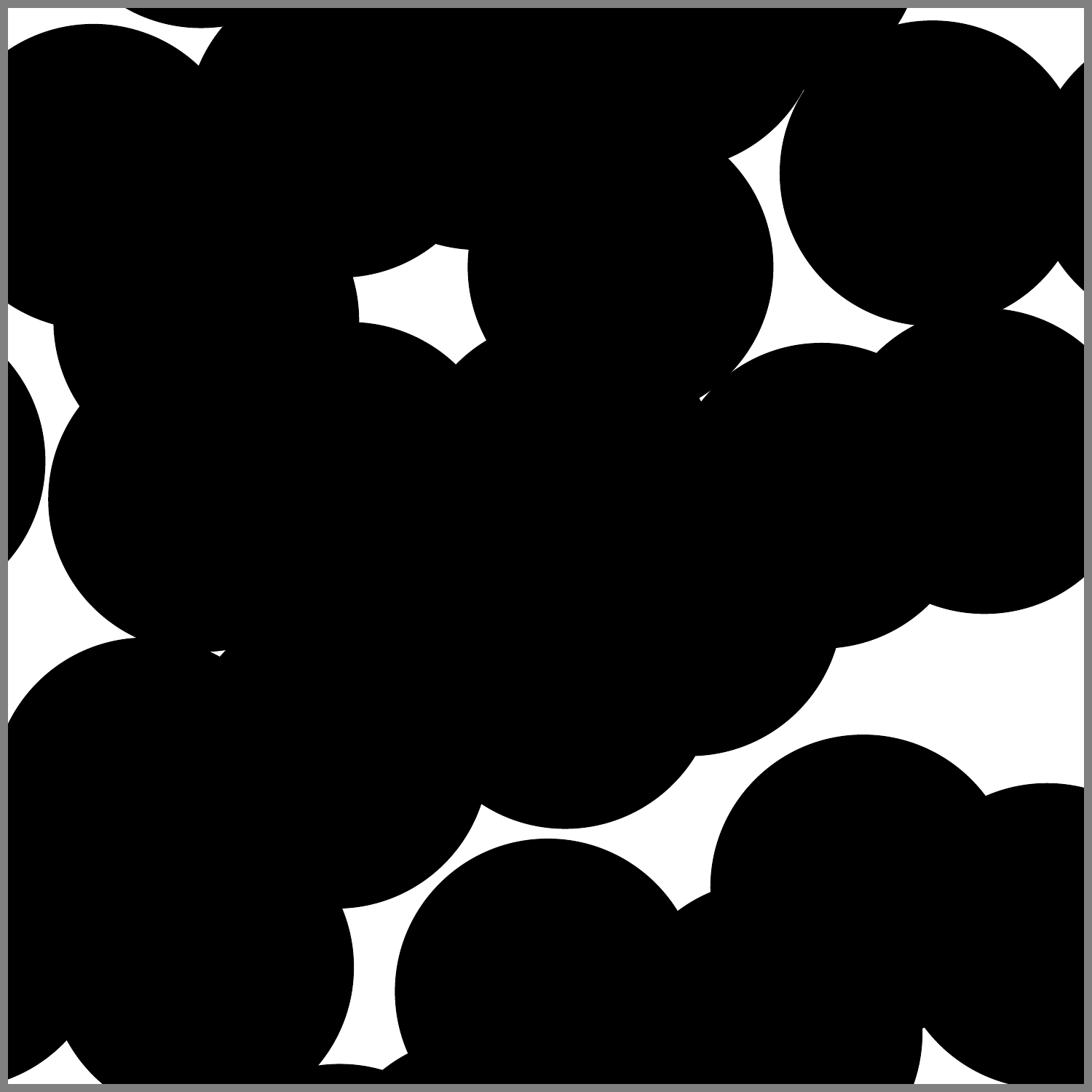}}\\
    \subfigure[][]{\includegraphics[width=0.49\linewidth]{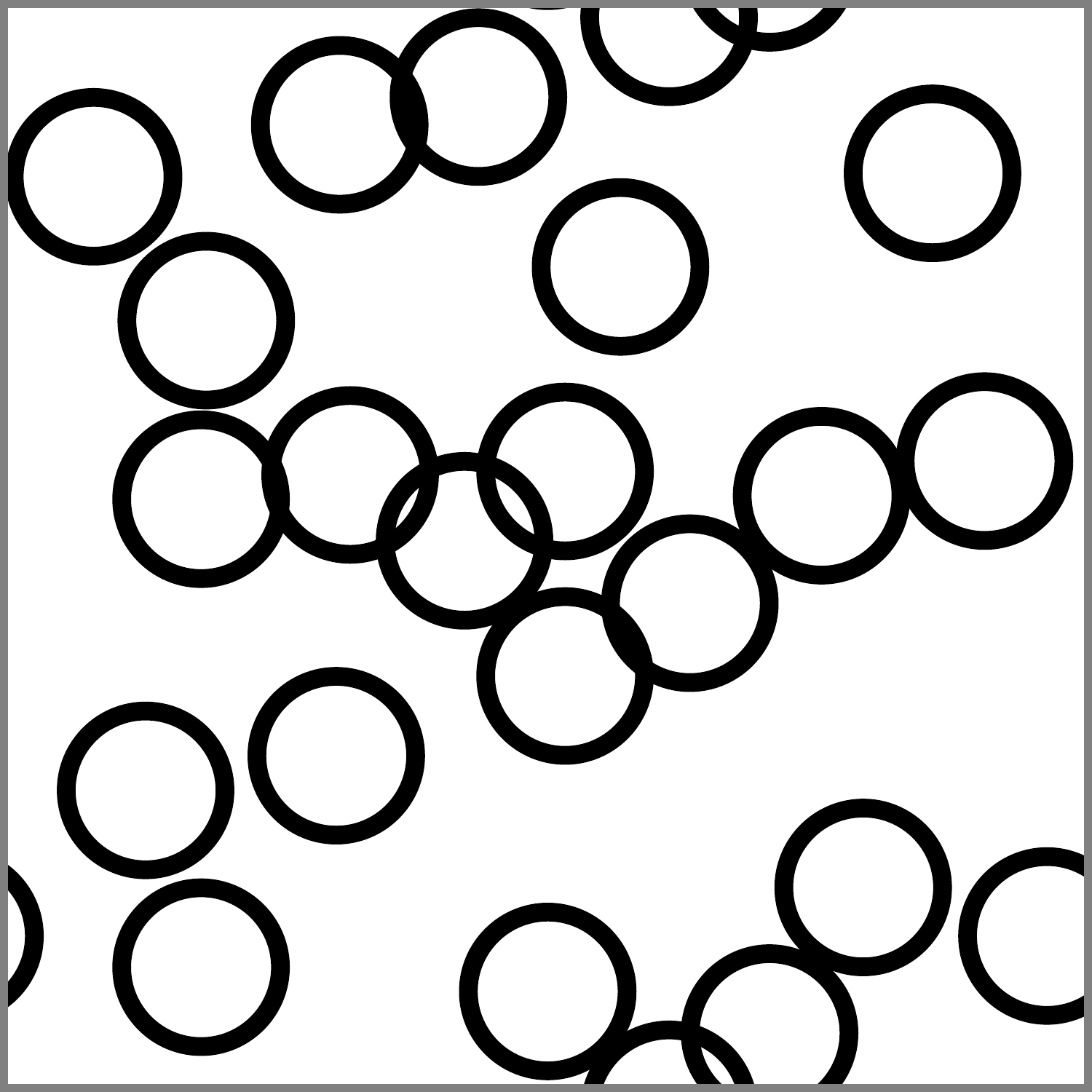}}
    \subfigure[][]{\includegraphics[width=0.49\linewidth]{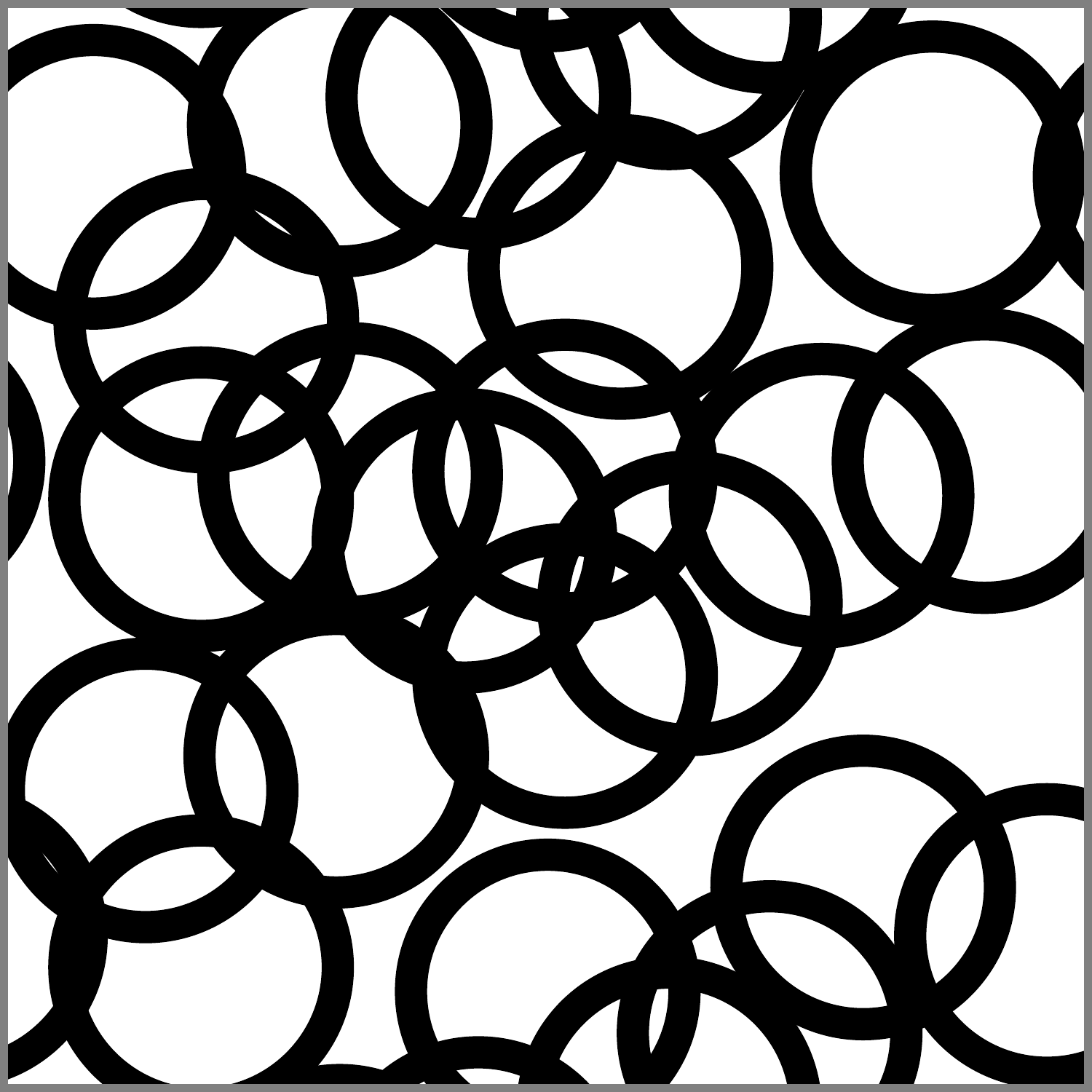}}
  \end{minipage}
  \caption{Morphometric analysis of a point pattern (left hand side):
  if to each point a disc of constant radius is assigned (a), the plane gets quickly covered by these discs even for intermediate radii (b).
  If instead a ring is assigned to each point (c), minute structural details can be resolved (d) even at large radii.}
  \label{fig:compare-discs-and-rings}
\end{figure}

The so-called Minkowski functionals from integral geometry~\cite{Santalo1976, SchneiderWeil2008, Mecke_Integralgeometrie_1994} provide a complementary approach, which rectifies the shortcomings of the standard $n$-point correlation function like the limitation to low order correlation functions \cite{mecke_robust_1994}.
The central idea is to assign to each point in the point pattern a ball of constant radius $r$, see Figs.~\ref{fig:compare-discs-and-rings}~(a). 
Thus, the point pattern is turned into a two-phase medium (or black and white image).
The latter can be characterized comprehensively, robustly, and efficiently by the Minkowski functionals~\cite{Mecke1998,MeckeOverview2000}.
The mean values of these Minkowski functionals as a function of the radius $r$ of the balls define a correlation function that implicitly contains information from $n$-point correlation functions of arbitrary order $n$~\cite{mecke_robust_1994,mecke_morphological_2005}.
Therefore, they can distinguish different point processes with the same structure factor.
Recent work has further established a close connection between germ–grain Minkowski functionals and $k$-nearest-neighbour distribution functions, showing that both approaches provide nearly identical constraints on cosmological parameters~\cite{Gangopadhyay2025}.
Minkowski functionals have already offered insights into astronomy~\cite{kerscher_minkowski_1997, kerscher_non-gaussian_2001, kerscher_morphological_2001, wiegand_direct_2014, Wiegand2017}, condensed matter physics~\cite{Jacobs_dewetting_1998,herminghaus_spinodal_1998}, and biology~\cite{mecke_morphological_2005}.
In~\cite{Voromink2026}, an algorithm is developed for estimating the Minkowski functionals and, more generally, the Minkowski tensors of a broad class of sets from point-cloud representations.

A major advantage of the Minkowski functionals is their additivity, that is, the functional value of the union of two bodies is given by the sum of their individual functional values minus that of their intersection.
Together with continuity under suitable geometric perturbations, this provides a mathematical basis for their robustness as shape descriptors.
%
Moreover, the Minkowski functionals contain the information of all continuous, motion-invariant, and additive scalar shape descriptors on convex bodies~\cite{Hadwiger1957}.
In this sense, the Minkowski functionals contain all robust multipoint information.

So far, however, these Minkowski correlation functions were limited to short distances, in the sense, that with a growing radius of the balls the whole system gets quickly covered by the spheres~\footnote{This is related to the so-called covering problem. The covering radius is the smallest radius at which the whole system is covered by the balls.}.
Therefore, the characteristic details of the point pattern can no longer be resolved, see Figs.~\ref{fig:compare-discs-and-rings}~(b).
This nearsightedness is a major disadvantage of the morphometric analysis for applications in physics, in particular for detecting large-scale structures in the Universe~\cite{mecke_robust_1994}, non-{Gaussian} morphology on large scales~\cite{kerscher_non-gaussian_2001}, long-range correlations in dewetting patterns~\cite{Jacobs_dewetting_1998}  or of tree positions in forests~\cite{mecke_morphological_2005}.  

Here, we lift this restriction by introducing ring-based Minkowski correlation functions (MCF) $M_j(r,s,\Phi)$ of point processes $\Phi$  that remain informative at long ranges.
The basic idea is to simply replace the discs by rings~\cite{Mecke_workingpaper_2008}, see Figs.~\ref{fig:compare-discs-and-rings}~(c) which was inspired by  a morphological model for suspensions of colloidal particles with wetting layers \cite{Mecke_fluids_1996,brodatzki_morphmodel_2001}.  
Unlike discs, rings do not necessarily cover the entire observation window as their outer radius increases, allowing the Minkowski functionals to remain sensitive to structural details at larger radii.
Thus, minute structure details can be resolved also at longer distances, see Figs.~\ref{fig:compare-discs-and-rings}~(d).

Similar to mark correlation functions~\cite{Beisbart_2002}  the Minkowski correlation functions $M_j(r,s,\Phi)$ is a method for  relating physical properties to spatial distributions $\Phi$ of points where inner radius $s$ and outer radius $r$  have  physical meaning such as wetting layers or a depletion zone of impact craters, for instance.

We outline how the MCFs can be defined in arbitrary dimensions $d$.
Then, we concentrate on planar point patterns from mathematical benchmark models over simulations to experimental data.
The MCFs are well defined for practically any point pattern, both in finite samples and in the thermodynamic limit.
We discuss their exact numerical evaluation and the correction of edge effects.
Moreover, we analytically relate the MCFs to two-point (and, more generally, $n$-point) information.
We show in the thin-ring limit how the pair-correlation function is recovered from the MCFs and illustrate higher-order sensitivity on experimental data.

The article is organized as follows.
In Sec.~\ref{Sec:MCF}, we introduce the Minkowski functionals, develop the geometric framework (positive reach) for unions of rings, and formally define the ring-based Minkowski correlation functions together with their series representation and thermodynamic limit.
In Sec.~\ref{Sec:Algorithm}, we present an exact and efficient algorithm for planar point patterns and discuss boundary corrections; we validate the implementation against closed-form results for the Poisson point process.
In Sec.~\ref{Sec:multipoint}, we analyze how ring thickness controls higher-order content, derive the thin-ring limit that recovers the pair-correlation function, and illustrate discrimination beyond second order using the Baddeley–Silverman process.
In Sec.~\ref{sec:applications}, we apply the method to three data sets across scales: active colloids (Quincke rollers), longleaf pine trees, and the Hubble Ultra Deep Field. The appendices collect technical bounds and proofs, including the positive-reach result and the Poisson (Miles) formulas.

\section{Minkowski correlation functions (MCF)}\label{Sec:MCF}

\begin{figure*}[t]
  \centering
  \subfigure[][]{\includegraphics[width=0.2\linewidth]{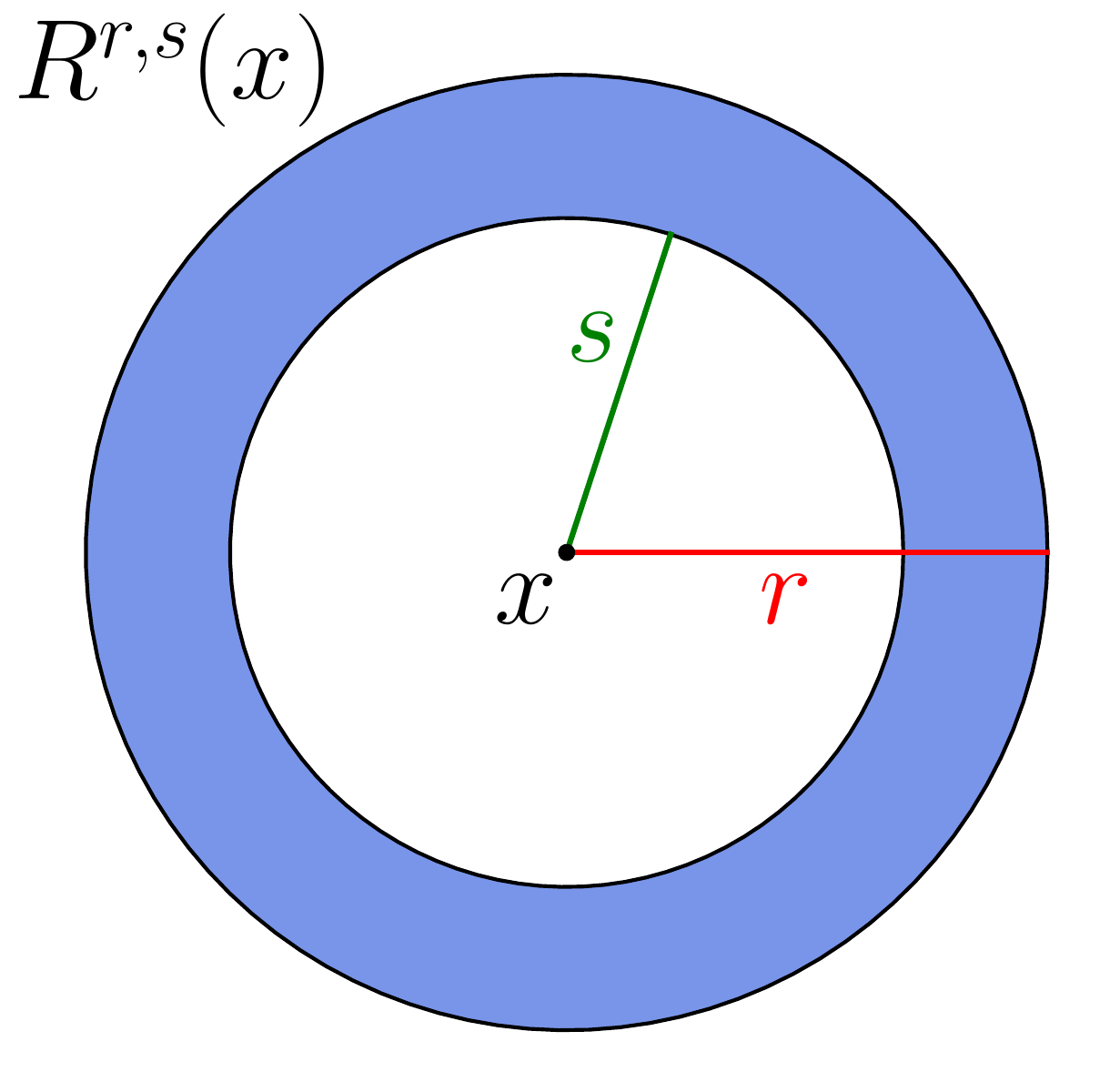}\label{fig:OneRing}}
  \subfigure[][\hbox{} Area]{\includegraphics[width=0.26\linewidth]{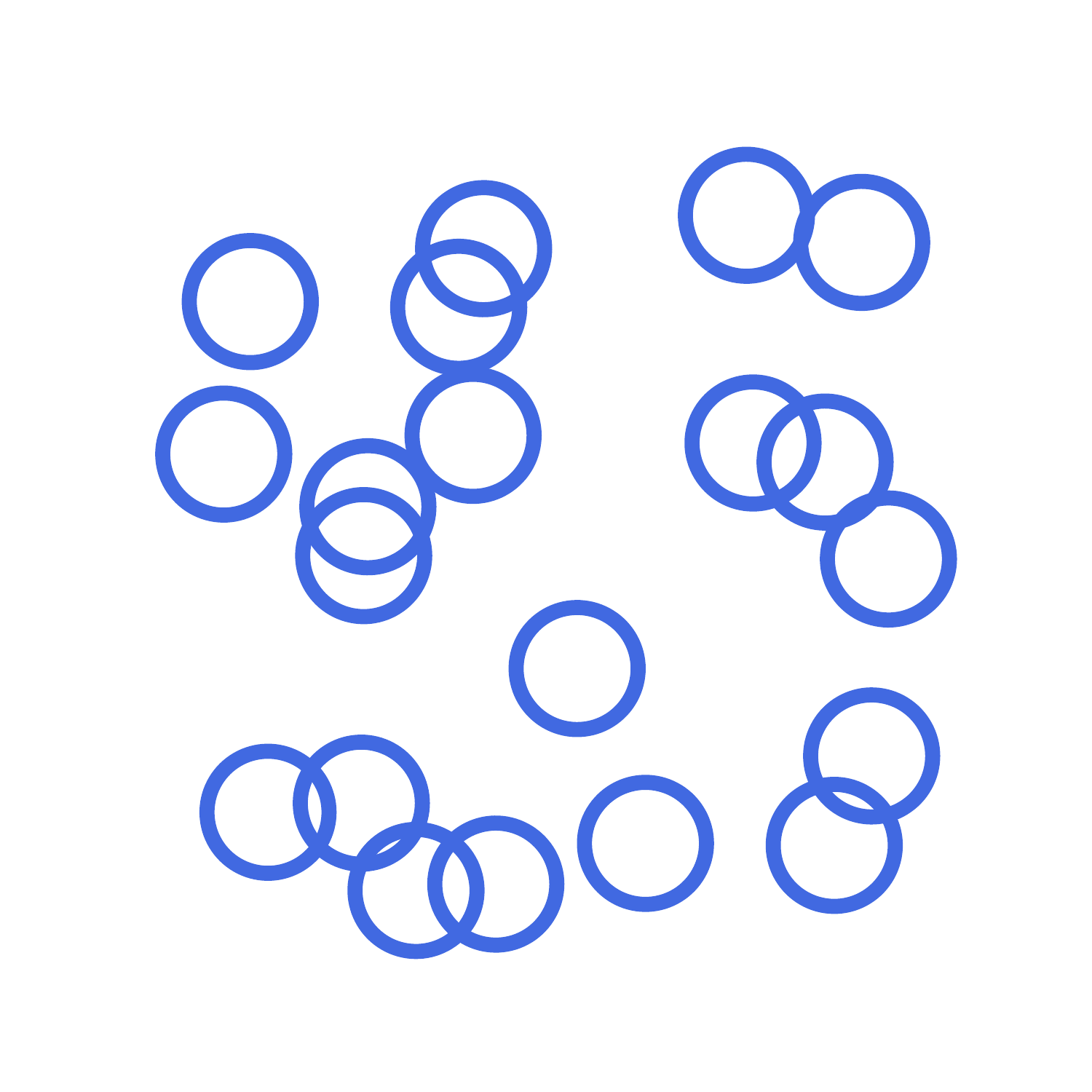}\label{fig:Rings_Volume}}
  \subfigure[][\hbox{} Perimeter]{\includegraphics[width=0.26\linewidth]{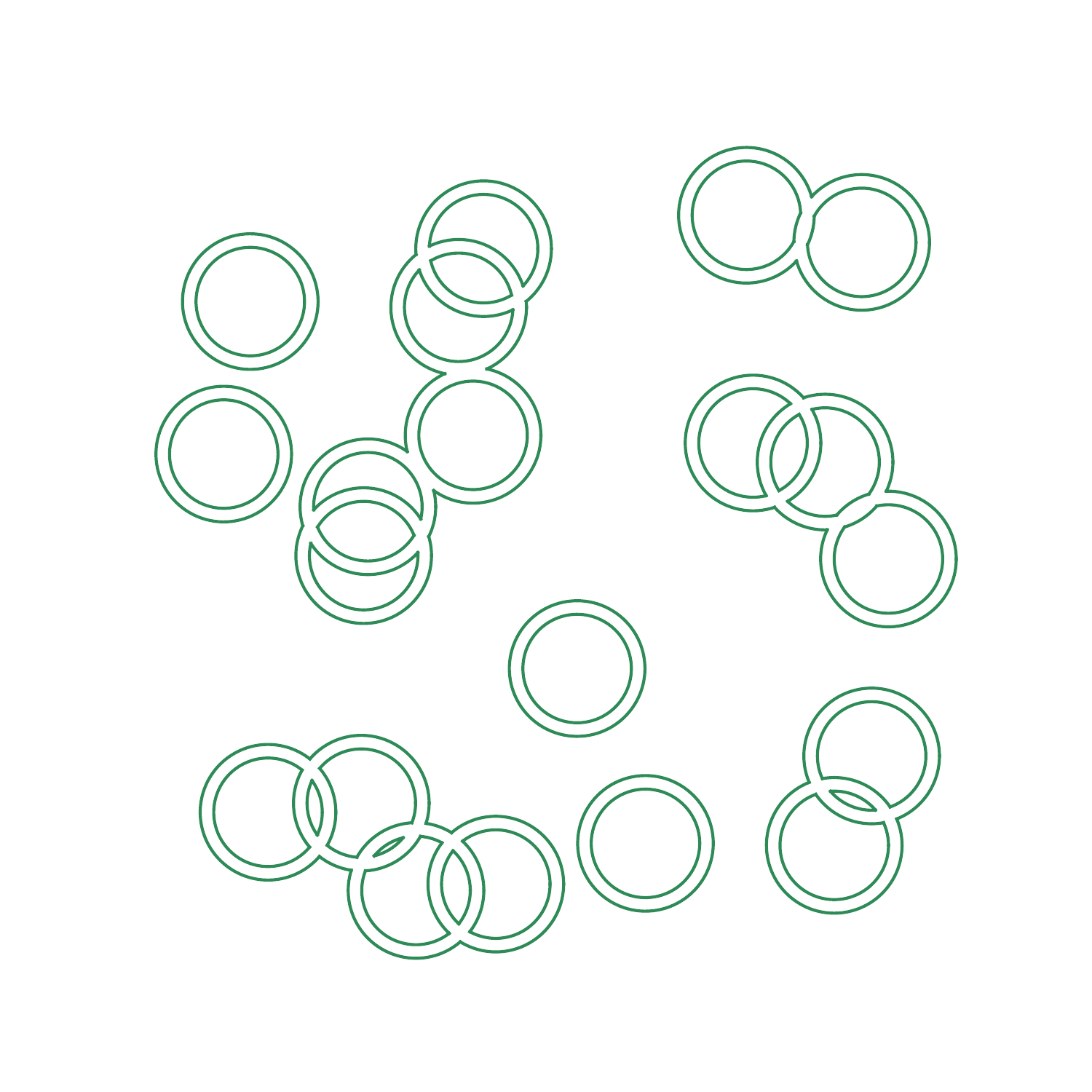}\label{fig:Rings_Surface}}
  \subfigure[][\hbox{} Euler characteristic]{\includegraphics[width=0.26\linewidth]{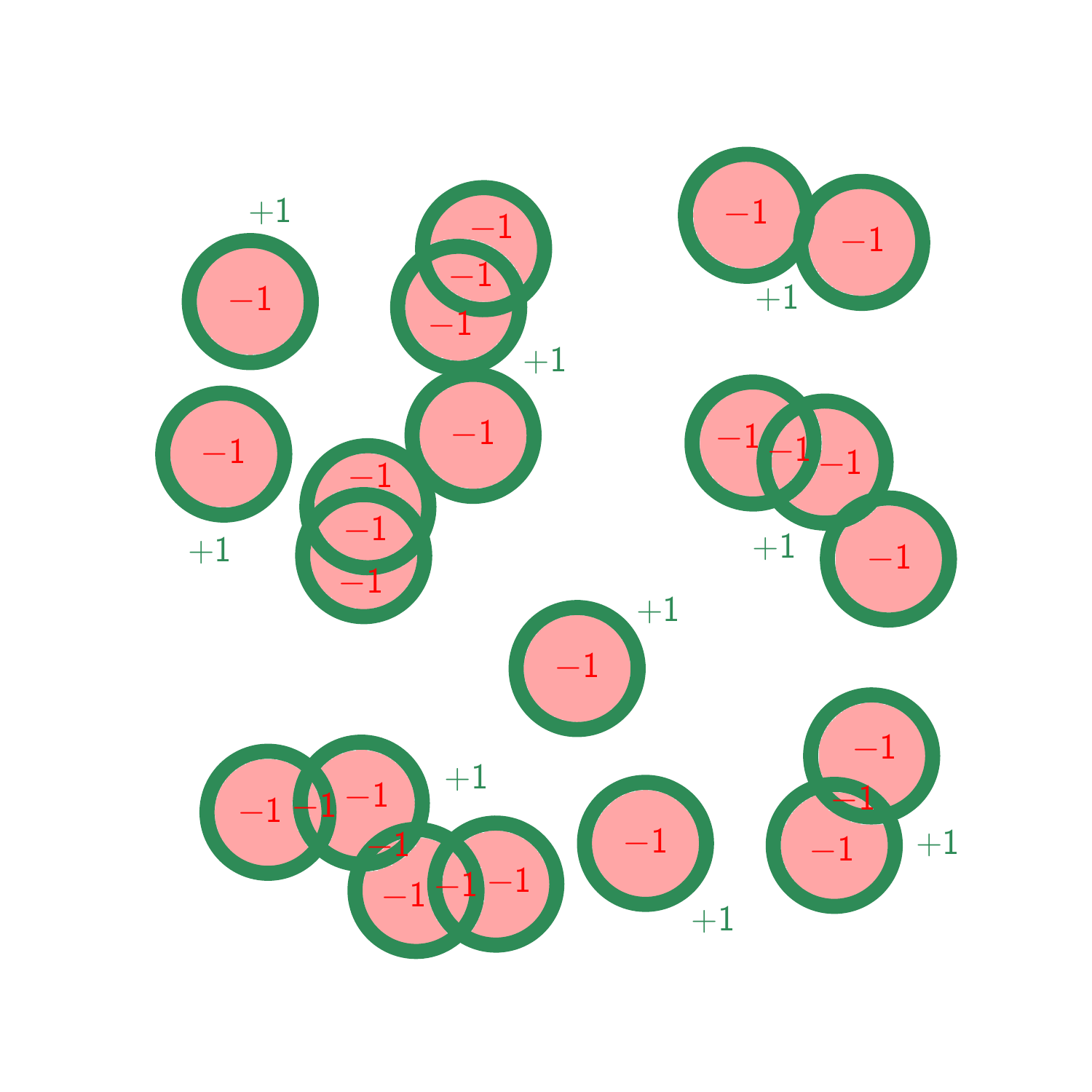}\label{fig:Rings_Euler}}
  \caption{Minkowski functionals of (overlapping) rings: (a) a ring is defined by its center $x$ and the outer and inner radii.
  The Minkowski functionals $M_{i}$ of the (overlapping) rings are in two dimensions given by the area, $M_0=A$ (b), half the perimeter, $M_1=\frac{1}{2}P$ (c), and the Euler characteristic, $M_2=\chi$ (d).
  The Euler characteristic $\chi$ is a topological constant that is given by the number of connected components minus the number of holes.}
  \label{fig:Minkowski-overlapping-rings}
\end{figure*}

We consider a stationary point process $\Phi$ in $\R^d$ observed in a finite observation window.
The definition of the Minkowski correlation functions (MCF) will be given in the thermodynamic limit, i.e., in the limit of an infinitely large observation window.
We assign to each point $x\in\Phi$ a spherical shell with outer radius $r$ and inner radius $s$
\begin{align*}
    R(x) := R^{r,s}(x) := \{ y\in\R^d \,:\, s\leq\Vert x-y\Vert\leq r\}.
\end{align*}
In two dimensions a spherical shell is a ring, see Fig.~\ref{fig:Minkowski-overlapping-rings}~(a).
The union $\cup_{x\in\Phi} R^{r,s}(x)$ defines a two-phase medium (foreground = rings, background = complement).
The central idea is to quantify the morphology $M_j(r,s,\Phi)$ of this two-phase medium as a function of the outer radius $r$ \cite{Mecke_workingpaper_2008}.

There are reasonable choices for the inner radius $s$.
If necessary, it can be used as a second argument of the MCF.
Often, it is sufficient to choose either a fixed ratio of $s$ and $r$ or of $s^2$ and $r^2$.
The first choice corresponds to a fixed ratio of the circumference of the outer and inner circle.
The second choice results in a fixed ratio of their areas.
In higher dimensions, a fixed ratio of, e.\,g., the volumes of the outer and inner spheres or of their surface area can be chosen.

\subsection{Minkowski functionals}\label{Sec:MinkowskiFunctionals}

The Minkowski functionals from integral geometry~\cite{SchneiderWeil2008} (see Section 14.2) provide a comprehensive and robust morphometric analysis, as explained in the introduction.
They are scalar descriptors that quantify the geometry of a two-phase medium in a $d$-dimensional space and map it to a real value, e.\,g., to its area.
Originally, the Minkowski functionals were defined and studied for convex sets and, subsequently, for finite unions of convex sets.
For the purposes of this work, however, this is not sufficient, since even a spherical shell does not belong to this class.
In \cite{Rataj}, the Minkowski functionals – there referred to as Lipschitz–Killing curvatures (Definition~4.31) – are introduced for more general sets.
We denote this class of sets, which includes convex bodies, by $\U$ and introduce it explicitly in Subsection~\ref{Sec:positiveReach}.

Let us first consider any real-valued functional $F$ on the set of convex bodies, that is at least continuous, motion invariant and additive.
The latter means that the functional value of the union of two bodies $K$ and $L$, which is convex, is given by the functional values of the single bodies and their intersection:
\begin{align}
  F(K\cup L) = F(K)+F(L)-F(K\cap L).
  \label{eq:additivity}
\end{align}
These conditions lead to a functional that is robust against noise, because a small deformation, heuristically, leads only to a small change in the functional value.

The importance of the Minkowski functionals as shape descriptors that quantify the geometry of a two-phase medium arises from the celebrated characterization theorem by Hadwiger~\cite{Hadwiger1957} for convex bodies and its generalizations.
Theorem 8.8 in \cite{Rataj} says that any such functional $F$ admits a unique motion invariant and additive extension to $\U$ and that this extension (also notated by $F$ here) can be written as a linear combination of the Minkowski functionals $M_0,\dots,M_d$
\begin{align}
  F(K) = \sum_{i=0}^{d} c_{i} M_{i}(K), \qquad K\in\U,
\end{align}
where $c_{i}$ are constants.
In this sense, the Minkowski functionals contain all additive shape information.

The most important special cases are the $d$-dimensional volume $M_0$, half the surface area $M_1$, and the Euler characteristic $M_d$.
The latter is provided by Theorem~4.53 in \cite{Rataj}, whereas the first two follow directly from Definition~4.31 in \cite{Rataj}.
Specifically, in dimension 2 we obtain the Minkowski functionals $M_0$ (area), $M_1$ (half the perimeter), and $M_2$ (Euler characteristic), see Fig.~\ref{fig:Minkowski-overlapping-rings}~(b)--(d).
The latter is a topological invariant, which, for a compact planar set, equals the number of components minus the number of holes, see Fig.~\ref{fig:Rings_Euler}.
The Minkowski functionals of a ring $R$ with outer radius $r$ and inner radius $s>0$, see Fig.~\ref{fig:OneRing}, are given by
\begin{align*}
      M_0(R) &= \pi(r^2-s^2) = A(R),\\
      M_1(R) &= \pi(r+s) = \frac{1}{2}P(R), \\
      M_2(R) &= 0 = \chi(R).
\end{align*}
In the case $s=0$, the formulas for $M_0(R),M_1(R)$ remain valid, but $M_2(R)$ takes the value 1.
For a more detailed introduction of the Minkowski functionals and exemplary applications, see~\cite{Mecke_Integralgeometrie_1994, Mecke1998, MeckeOverview2000, SchroederTurketal2011AdvMater}.
They have already been successfully applied to a variety of physical systems~\cite{Mecke1998, MeckeStoyan2000}, in condensed and soft matter~\cite{MeckeStoyan2002, HansenGoosMecke2009, KapferEtAl2012, WittmannEtAl2014, ScholzWirnerEtAl2015}, 
as well as in astronomy and cosmology~\cite{kerscher_morphological_2001, Schmalzing1999, GoeringKlattEtAl2013} or in pattern and porous analysis~\cite{Mecke1996, Becker:2003, MantzJacobsMecke2008, Arns_BooleanReconstruction_2003, Arns_BooleanReconstruction_2009, ScholzEtAl2015}.

For typical porous materials, the Minkowski functionals grow on average proportional to the volume of the observation window $W$.
Therefore, we rescale the Minkowski functionals by the volume $|W|$ of the observation window so that they converge to a well-defined value in the thermodynamic limit ($|W|\rightarrow \infty$).
We will show that this limit exists and does not depend on the sequence $(W_n)_{n\in\N}$ of observation windows as long as the sets $W_n$ are convex such that the inradius converges to $\infty$.

\subsection{Positive reach of intersections of spherical shells}\label{Sec:positiveReach}

In this section, we will present a result that shows that a union of spherical shells belongs, for almost all centers, to the class $\U$, which we will define explicitly below.
The proof of the result can be found in Appendix~\ref{App:proof}.
Essential for this is the notion of sets with positive reach.

A set $A$ is said to have reach $t$, if $t$ is the largest number satisfying the following property: for every point $x$ whose distance to $A$ is less than $t$, the closest point in $A$ to $x$ is unique.
The reach of $A$ equals $0$ if no $t>0$ satisfies this property, and $\infty$ if every $t>0$ does.
We say that $A$ has positive reach if $\reach(A)>0$.
Observe that a set with positive reach is closed and the reach is $\infty$ if and only if the set is convex and closed.
For a precise mathematical definition, we refer to \cite{Rataj} (see Chapter 4).
The reach of a ring $R^{r,s}$ with outer radius $r$ and inner radius $s$ is given by
\begin{align*}
    \reach(R^{r,s}) = \begin{cases}
    s, & s>0, \\
    \infty, & s=0.
    \end{cases}
\end{align*}
The class $\U$ consists of all sets that can be written as a finite union of compact sets $A_1, \dots, A_k$ with positive reach, such that every intersection $\cap_{i \in I} A_i$, $I \subseteq \{1, \dots, k\}$, also has positive reach.
For such sets, the Minkowski functionals, which are defined for sets of positive reach, can be determined by additive extension.

\begin{Th} \label{Th:positiveReachIntersections}
Fix $0<s<r$.
For any $n\in\mathbb{N}$ and for almost all $x_1,\dots,x_n\in\R^d$, the intersection $\cap_{i=1}^n R^{r,s}(x_i)$ has positive reach.
Consequently, the finite union $\cup_{i=1}^n R^{r,s}(x_i)$ belongs to $\U$.
\end{Th}

Note that Theorem~\ref{Th:positiveReachIntersections} ensures that, for almost all $x_1, \dots, x_n \in \R^d$, the union $\cup_{i=1}^n R(x_i)$ belongs to the class $\U$, for which the Minkowski functionals are available.

\subsection{Definition of the MCFs}\label{Sec:DefinitionMCF}

From now on we consider a stationary point process $\Phi$ on $\R^d$ with existing correlation functions $\rho_n:(\R^d)^n\rightarrow [0,\infty)$ for all $n\in\N$, that is for all $n\in\N$ the equality
\begin{align*}
    &\E{\sum_{x_1,\dots,x_n\in \Phi}^{\neq} f(x_1,\dots,x_n)} \\
    &\hspace{20pt}=\, \int_{(\R^d)^n} f(x_1,\dots,x_n)\rho_n(x_1,\dots,x_n) \;\d{(x_1,\dots,x_n)}
\end{align*}
holds for all $f:(\R^d)^n\rightarrow [0,\infty)$, where $\sum^{\neq}$ indicates that the sum is taken over pairwise different points of $\Phi$.
The stationarity of $\Phi$ implies that the functions $\rho_n$ are translation invariant, that is
\begin{align*}
    \rho_n(x_1+t,\dots,x_n+t) \,=\, \rho_n(x_1,\dots,x_n)
\end{align*}
for all $t,x_1,\dots,x_n\in\R^d$.
From this it follows, for example, that the one-point correlation function $\rho_1$ is constant.
This constant value $\gamma$ is called the intensity and equals the mean number of points per unit $d$-dimensional volume.
The pair-correlation function is $g^{(2)}(x,y) := \rho_2(x,y)\gamma^{-2}$; for stationary and isotropic processes $g^{(2)}$ depends only on the distance.
It is the real-space analogue of the structure factor.
We assume that there is a constant $C>0$ such that
\begin{align} \label{Korfkt:uniformlybounded}
        \rho_n(x_1,\dots,x_n) \,\leq\, C^n
\end{align}
for all $n\in\N$ and almost all $x_1,\dots,x_n\in\R^d$.
Then
\begin{align}\label{MCF-definition-W}
    M_j(r,s, \Phi,W) \,:=\, \E{ M_j\Big( \bigcup_{x \in \Phi \cap W} R^{r,s}(x) \Big) }
\end{align}
denotes the $j$-th Minkowski functional evaluated on the union of rings centered at the points of $\Phi$ within $W$.
By Theorem~\ref{Th:positiveReachIntersections}, the expression within the expected value is almost surely uniquely defined.
Since the definition of the Minkowski correlation functionals should not depend on the choice of an observation window, we will take the thermodynamic limit in which the volume of the observation window approaches infinity.
We will see that, under certain conditions on the sequence of observation windows, this limit is independent of the particular sequence chosen.
To mathematically prove the existence of the limit and to derive a formula for its representation, we will focus here on the two-dimensional case $d=2$ and note that the argument applies to the volume $M_0$ and the surface area $M_1$ in all dimensions. \\
Let $(W_m)_{m \in \mathbb{N}}$ be a sequence of compact convex sets such that the inradius $r(W_m)$ tends to infinity.
The inradius of a convex set $A$ is defined as the radius of the largest ball contained in $A$.
For example, $W_m$ could be a cube with a side length of $m$ or a ball with a radius of $m$.
We define the $j$-th Minkowski correlation function (MCF) of $\Phi$ by
\begin{align}\label{MCF-definition} 
    M_j(r,s,\Phi) \,:=\, \lim_{m\rightarrow\infty} \frac{M_j(r,s,\Phi,W_m)}{|W_m|}.
\end{align}

\begin{Prop}[Series representation]\label{prop:MCF_series}
  Let $0<s<r$ and assume either $j\in\{0,1\}$ or $d=j=2$.
  Assume the product densities satisfy the bound \eqref{Korfkt:uniformlybounded} and let $(W_m)$ be a sequence of compact convex sets with $r(W_m)\rightarrow\infty$.
  Then the limit \eqref{MCF-definition} exists and
  \begin{align} \label{eq:Formula_MCF}
    M_j(r,s,\Phi) \,=\, \sum_{n=1}^\infty \frac{(-1)^{n+1}}{n!} g_j(r,s,\Phi,n),
\end{align}
where $g_j(r,s,\Phi,n)$ is given by the integral
\begin{align} \label{eq:limit_integrals}
    \int_{(\R^d)^{n-1}} f(\0,x_2,\dots,x_n) \;\d{(x_2,\dots,x_{n})}.
\end{align}
with the function $f:(\R^d)^n\rightarrow\R$ given by
\begin{align*}
    f(x_1,\dots,x_n) \,:=\, M_j\left( \cap_{i=1}^{n} R(x_i) \right) \rho_n(x_1,\dots,x_{n}).
\end{align*}
\end{Prop}

This follows from \eqref{eq:Formula_MCF_appendix}.
In particular, the value of $M_j(r,s,\Phi)$ does not depend on the specific choice of the sequence $(W_m)_{m \in \mathbb{N}}$ of observation windows.
If only a single sample within an observation window is given, its Minkowski functionals normalized by the volume of the window can serve as an estimate of the mean value.
The functional values are global properties and correspond to a spatial average.
Note that for an ergodic system, the spatial average coincides in the thermodynamic limit with the ensemble average. \\
For $s=0$, MCFs reduce to the previously studied correlation functions for overlapping disks/balls.
For $s>0$, MCFs retain sensitivity to higher-order correlations at radii where the disk-based functions have already saturated.
In Sec.~\ref{Sec:multipoint}, we show that the limit $s\uparrow r$ recovers the pair-correlation function $g^{(2)}$.

\section{Exact numerical algorithm}\label{Sec:Algorithm}

Here, we present explicit formulae of the Minkowski functionals of overlapping rings for any planar point pattern (as a function of the coordinates of the points). Corresponding formulas for an ensemble of overlapping discs can be found in~\cite{brodatzki_simulating_2002} and for rings in~\cite{Brodatzki_diss_2003}. 
First algorithms for additive correlation functions have been developed in \cite{Brodatzki_diss_2003,Bittner_diploma_2008} and already used in a morphological model for complex fluids, where wetting layers of hard colloidal particles requires the calculation of ring-based structures \cite{Mecke_fluids_1996,brodatzki_morphmodel_2001}.

\subsection{Explicit formulae by summing local contributions}\label{ExplicitFormulae}

For the implementation of an efficient algorithm to compute the Minkowski functionals $M_0, M_1, M_2$ for a union of rings, we make use of the fact that the Minkowski functionals can be expressed as integrals over the boundary.
An explicit formula in arbitrary dimension for the Minkowski functionals of a compact set with $C^2$ boundary, given in terms of an integral over the boundary, can be found in \cite{Mecke_Integralgeometrie_1994, MeckeOverview2000} (Section~2).
For a union of finitely many rings $A$, however, these formulas are not directly applicable, since discontinuities occur at the intersections of the boundaries of the individual rings.
Nevertheless, suitable variants of these formulas exist that can be applied in this setting.
For the (half) perimeter $M_1$, this is trivial
\begin{align*}
    M_1(A) = \frac{1}{2} \int_{\partial A} \; \mathcal{H}(\d x),
\end{align*}
where $\mathcal{H}$ denotes the $(d-1)$-dimensional Hausdorff measure.
Denoting the unit normal vector at $x\in\partial A$ by $n(x)$ (which is almost everywhere defined), the divergence theorem yields for the area
\begin{align*}
    M_0(A) \,=\, \frac{1}{2} \int_{\partial A} \langle x,n(x)\rangle \; \mathcal{H}(\d x).
\end{align*}
Finally by the Gauss--Bonnet theorem, we obtain
\begin{align*}
    M_2(A) = \frac{1}{2\pi}\int_{\partial A} \kappa(x) \;\mathcal{H}(\d x) + \frac{1}{2\pi}\sum_{i=1}^k \theta_i,
\end{align*}
where $\kappa(x)$ is the geodesic curvature at $x\in\partial A$ and $\theta_1,\dots,\theta_k$ are the exterior angles at the points of discontinuity of the boundary.
All of these formulas, which apply to more general sets, can be evaluated explicitly in the present case.
To this end, we first observe that the boundary of $A$ can be represented as a union of circular arcs that intersect only at their endpoints.
Each arc lies either on an outer circle of radius $r$ or on an inner circle of radius $s$.
We denote by $\gamma_1,\dots,\gamma_m$ those arcs on outer circles and by $\rho_{m+1},\dots,\rho_{m+\ell}$ those on inner circles.
Writing $(x_i, y_i)$, $\alpha_i$, and $\beta_i$ for the center, initial angle, and terminal angle of the $i$-th arc, respectively, the above formulas yield
\begin{align*}
    M_0(A) &= \frac{1}{2} \sum_{i=1}^m g(x_i,y_i,r,\alpha_i,\beta_i) \\
    &\phantom{=} -\frac{1}{2} \sum_{i=m+1}^{m+\ell} g(x_i,y_i,s,\alpha_i,\beta_i), \\
    M_1(A) &= \frac{1}{2} \left( \sum_{i=1}^m \mathcal{H}(\gamma_i) + \sum_{i=m+1}^{m+\ell} \mathcal{H}(\rho_i) \right), \\
    M_2(A) &= \frac{1}{2\pi} \left( \frac{1}{r} \sum_{i=1}^m \mathcal{H}(\gamma_i) - \frac{1}{s} \sum_{i=m+1}^{m+\ell} \mathcal{H}(\rho_i) \right) + \frac{1}{2\pi}\sum_{i=1}^k \theta_i.
\end{align*}
where $g(x,y,t,\alpha,\beta)$ stands for
\begin{align*}
    t\left[ x(\sin(\beta)-\sin(\alpha)) + y(\cos(\alpha)-\cos(\beta)) + t(\beta-\alpha) \right].
\end{align*}

This enables us to develop an algorithm that is both efficient and precise.

\begin{Alg}
Compute the Minkowski functionals of a union of rings with centers $x_1, \dots, x_n \in \R^2$.
\begin{enumerate}
    \item[$\mathrm{(1)}$] Identify the boundary of the union $\bigcup_{i=1}^n R(x_i)$, represented by a collection of arcs $\gamma_1, \dots, \gamma_\ell$, and determine the exterior angles $\theta_1, \dots, \theta_k$ at the points of discontinuity of the boundary.
    \item[$\mathrm{(2)}$] Evaluate the prescribed integrals over each arc $\gamma_i$ and apply the formulas above to compute the Minkowski functionals.
\end{enumerate}
\end{Alg}

Determining the boundary in Step~(1) is the most challenging and computationally expensive part of the algorithm.
The method is mathematically rigorous and yields exact results in theory; numerical inaccuracies arise only from finite-precision arithmetic.
The algorithm is implemented in Python, and the code is publicly
available as the open-source package \texttt{minkowski-correlations}
~\cite{pabst_ZenodoCode}.

\subsection{Boundary conditions to correct edge effects}\label{Sec:BoundaryConditions}

For points $x_1,\dots,x_n$ in a finite observation, the rings can intersect the boundary of the window and rings of points lying outside the window might extend into the window.
Another advantage of the Minkowski functionals is that such edge effects due to the finite system size can easily be corrected by choosing appropriate boundary conditions for the Minkowski functionals.
There are several ways to address this, and our algorithm provides the following three.
The last two methods require the observation window to be a (axis-aligned) rectangle.
To correctly compute the volume $M_0$, the part of the observation window's boundary covered by the rings is also taken into account.
By default, such contributions are not included in the computation of $M_1$ and $M_2$; an option to include them is provided (see minus sampling below).

\begin{figure*}[t]
  \centering
  \subfigure[][]{\includegraphics[width=0.32\linewidth]{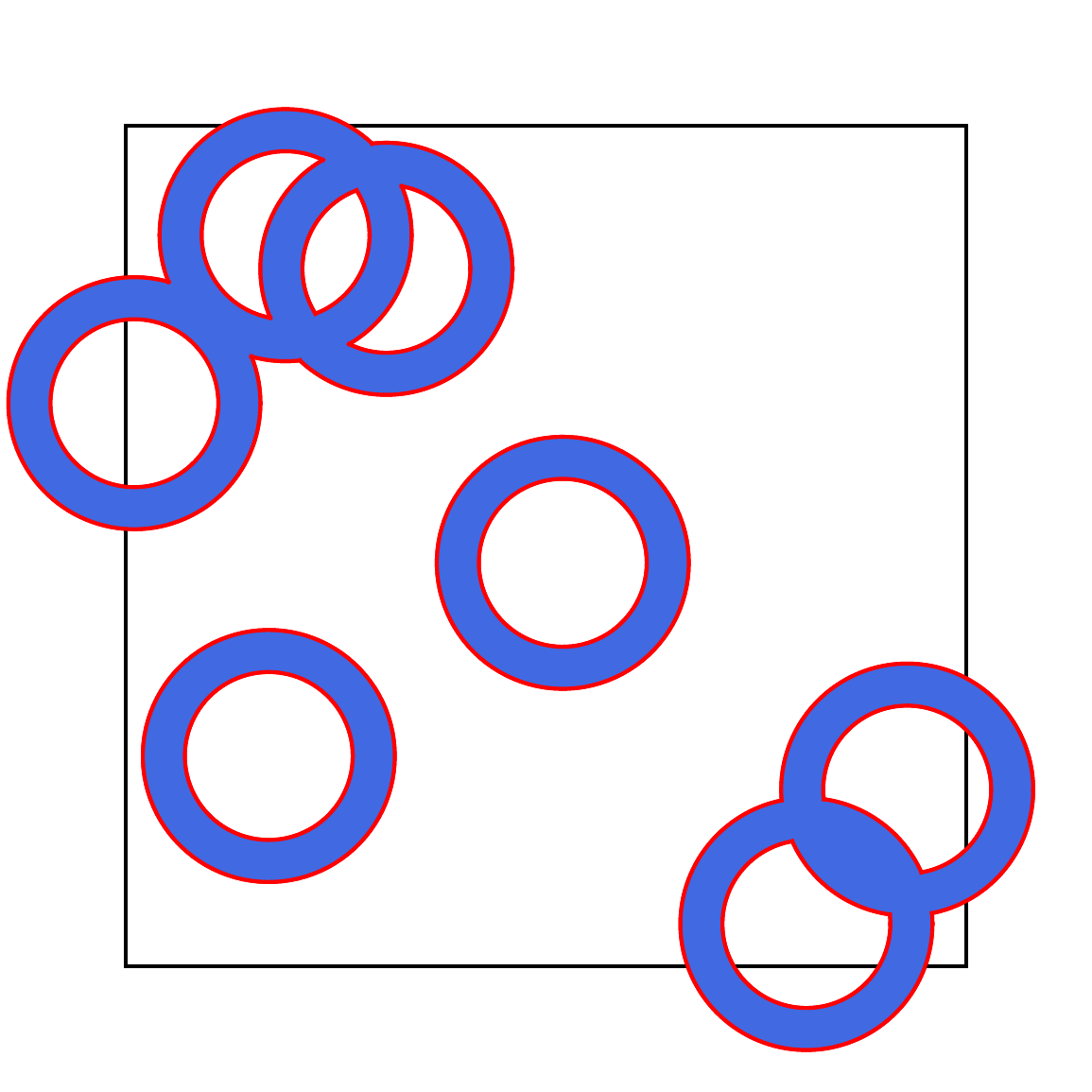}\label{fig:Boundary_None}}
  \subfigure[][]{\includegraphics[width=0.32\linewidth]{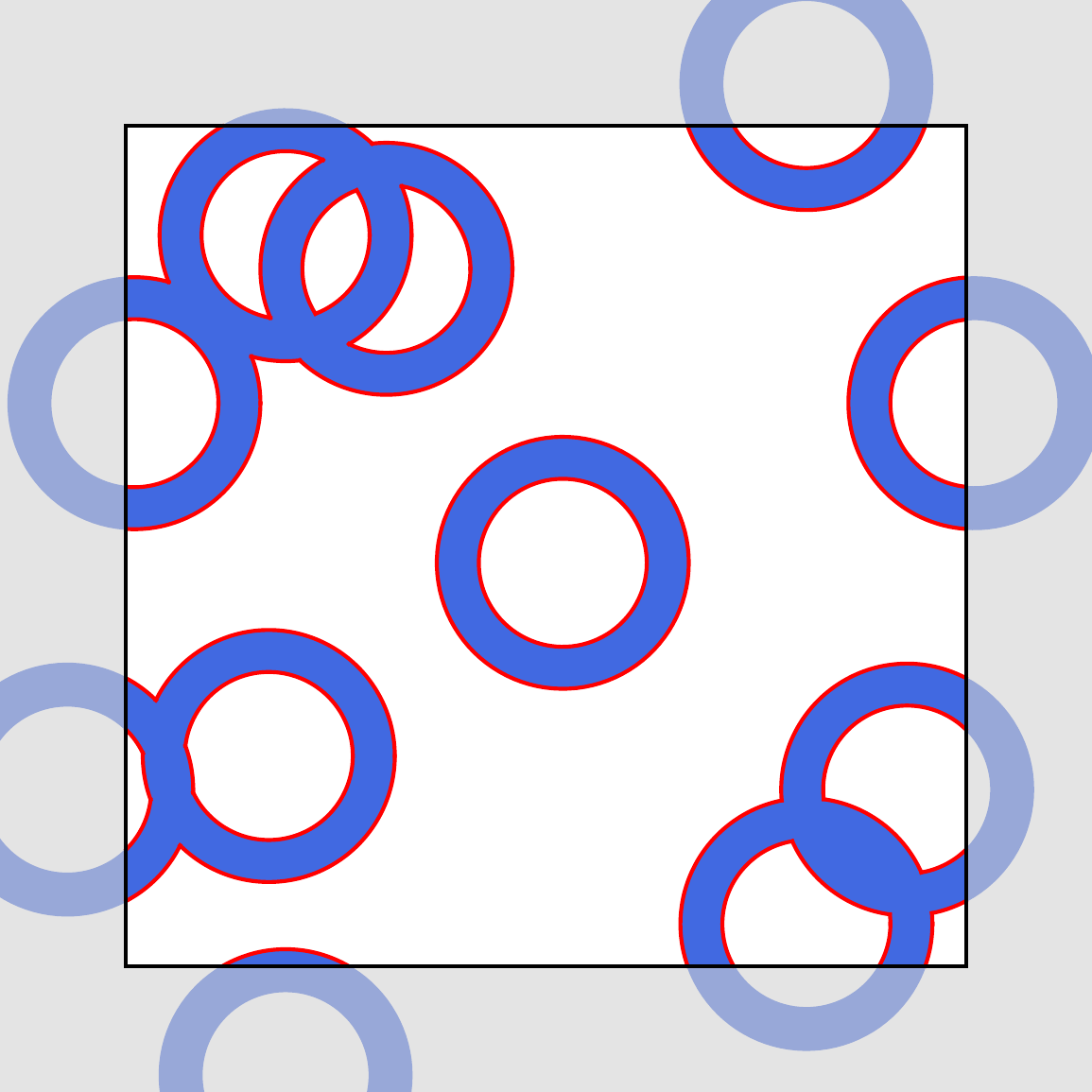}\label{fig:Boundary_Periodic}}
  \subfigure[][]{\includegraphics[width=0.32\linewidth]{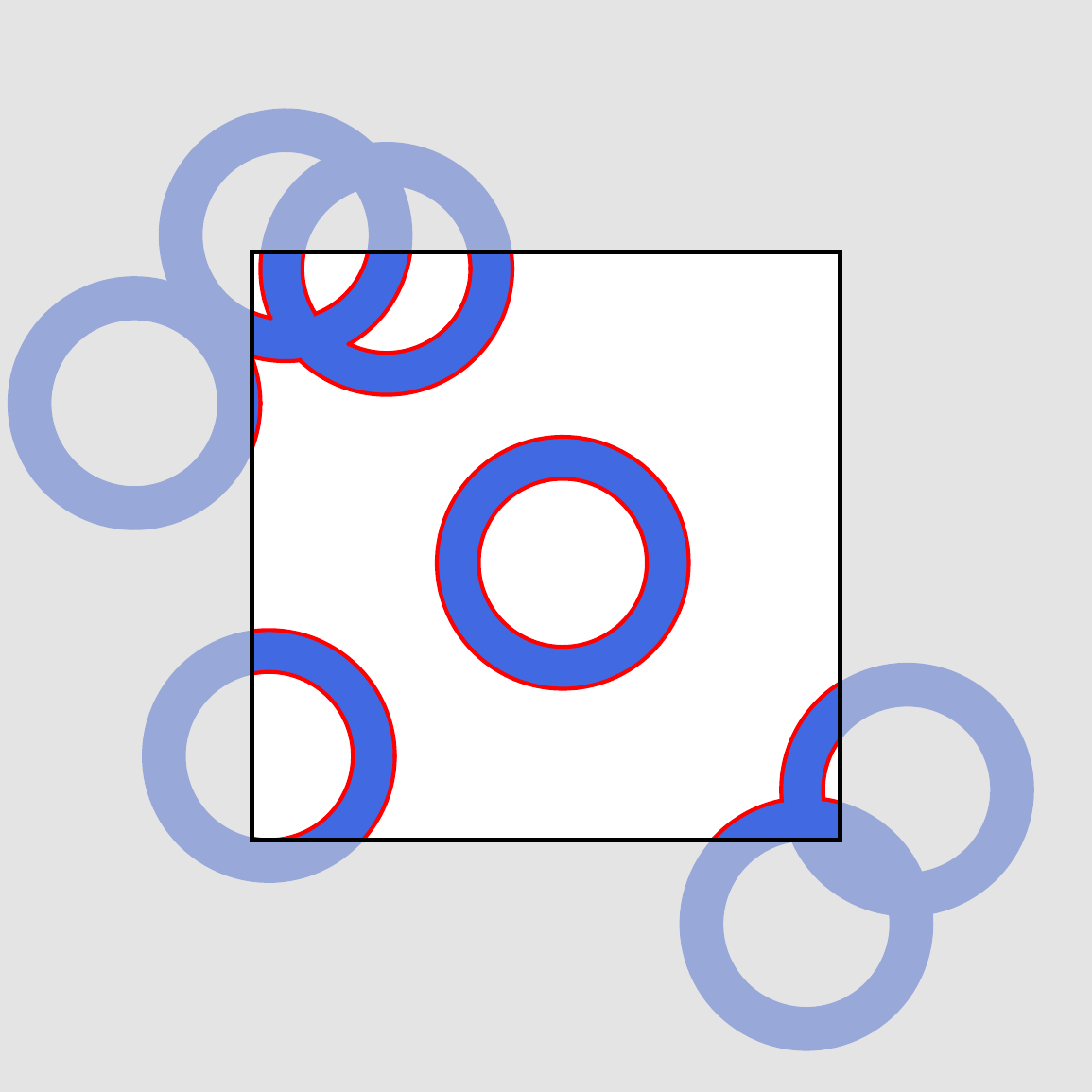}\label{fig:Boundary_MinusSampling}}
  \caption{A union of rings (a) without boundary conditions (b) with periodic boundary conditions and (c) using minus sampling with the boundary used by the algorithm highlighted in red in each figure.}
  \label{fig:BoundaryConditions}
\end{figure*}

A straightforward approach is to ignore the window and just compute the Minkowski functionals of the union $R(x_1)\cup\dots\cup R(x_n)$.
In this case, only the coordinates of the points matter, not the observation window in which they were observed (see Figure~\ref{fig:Boundary_None}).

An alternative is to impose periodic boundary conditions (Figure~\ref{fig:Boundary_Periodic}), under which rings extending beyond one side of the observation window re-emerge from the opposite side.
Periodic boundary conditions assume that the observation window is the unit cell of a periodic crystal, that is, an object leaving the left hand side of the observation window enters it on the right hand side.
They are especially useful to reduce edge effects.
For a homogeneous Poisson process, periodic boundary conditions do not introduce systematic bias in correlation functions.

Finally, there is the option of shrinking the observation window and treating the points that fall outside as boundary conditions.
This reduction, called minus sampling (Figure~\ref{fig:Boundary_MinusSampling}), is performed by subtracting the outer radius $r$ from each side of the window, which is only possible if both side lengths are greater than $2r$.
At the cost of effectively reducing the statistics, these boundary conditions avoid boundary-induced bias.
The algorithm provides the option to also include the part of the observation window boundary covered by rings in the computation of $M_1$ and $M_2$.
As a result, the estimated value of $M_2$ is indeed an integer, which is generally not the case when using minus sampling.

\subsection{The uncorrelated Poisson point process as benchmark test}\label{Sec:PoissonProcess}

To demonstrate the validity of the numerical algorithm, we compare the numerical to analytic results for the grand-canonical ensemble of the ideal gas, which is also a benchmark model of stochastic geometry called the Poisson point process.
The points in the point process are completely independent from each other and distributed randomly and uniformly in space~\cite{daley_introduction_2003, LastPenrose16}.
A stationary Poisson process with intensity $\gamma>0$ can be simulated within a bounded observation window $W$ by first drawing a Poisson-distributed random variable $\tau$ with parameter $\gamma |W|$, and then generating $\tau$ independent random points $X_1,\dots,X_\tau$ uniformly distributed in $W$.
Proposition 3.5 in \cite{LastPenrose16} shows that the resulting point pattern has the distribution of a stationary Poisson process with intensity $\gamma$ restricted to $W$.

The overlapping rings then correspond to the so-called Boolean model~\cite{Stoyan_Boolean_2005} which is also a standard model for porous materials~\cite{Mecke_SpatialPattern_1997,MeckeArns_2005,Arns_BooleanReconstruction_2003,Arns_BooleanReconstruction_2009,ScholzWirnerEtAl2015}.
For Boolean models with convex grains, the first and second moments of the Minkowski functionals are in many cases analytically known as well as a central limit theorem for their probability distribution~\cite{MeckeWagner1991, MeckeOverview2000, Mecke2001, kerscher_morphological_2001, HugLastSchulte2013, hug_second_2016}.
Such results may provide a basis for developing hypothesis tests based on the MCFs.
A rejection of the hypothesis of complete spatial randomness would indicate a deviation from the Poisson model and motivate further investigation.
For a stationary Poisson process $\Phi$ with intensity $\gamma$, the Minkowski functionals $M_0(r,s,\Phi)$ and $M_1(r,s,\Phi)$ can be computed in arbitrary dimension directly from \eqref{eq:Formula_MCF}.  
To this end, observe that the correlation functions of $\Phi$ are given by $\rho_n \equiv \gamma^n$.  
This follows, for example, from the multivariate Mecke equation (Theorem~4.4 in \cite{LastPenrose16}).
The computation of the integrals in \eqref{eq:limit_integrals} is based on the proof of Theorem 5.2.1 in \cite{SchneiderWeil2008}.

\begin{Lem}[Miles’ formulas] \label{Lem:Miles}
    For a stationary Poisson process $\Phi$ on $\R^d$ with intensity $\gamma$ we have
    \begin{align*}
        M_0(r,s,\Phi) \,&=\, 1-\exp\left( -\gamma M_0(R^{r,s}(\0)) \right), \\
        M_{1}(r,s,\Phi) \,&=\, \gamma M_{1}(R^{r,s}(\0)) \exp\left( -\gamma M_0(R^{r,s}(\0)) \right).
    \end{align*}
\end{Lem}

The formulas from Lemma~\ref{Lem:Miles} agree with the celebrated Miles formulas, even though the Miles formulas refer to the limit in a different asymptotic scenario.
Here, we obtain these formulas for ring-shaped grains.
For convex sets, they can be found in the isotropic case, for instance in \cite{SchneiderWeil2008} (Theorem 9.1.4).
We strongly believe that the Miles formulas also hold in full generality for rings, and therefore we use the Miles formula for the Euler characteristic $M_d$ in $d=2$ to compare it with our simulation results.
As we will see, the results strongly support this conjecture.
In two dimensions, the Miles formula for the Euler characteristic gives
\begin{align} \label{Miles_Euler}
    \widetilde{M}_2(\Phi) \,:=\, -\frac{\gamma^2}{\pi} M_1(R(\0))^2 \exp\left( -\gamma M_0(R(\0)) \right),
\end{align}
where we use a different notation here, since this is not a proven formula for $M_2$.
We note that the formula differs in the case of balls, i.e.\ for $s=0$, since an additional term appears which vanishes for $s>0$ because $M_2(R(\mathbf{0}))=0$.
By inserting the specific values of the Minkowski functionals into the formulas of Lemma~\ref{Lem:Miles} and into~\eqref{Miles_Euler}, we arrive at
\begin{align*}
    M_0(r,s,\Phi) \,&=\, 1-\exp\left( -\gamma\pi(r^2-s^2) \right), \\
    M_1(r,s,\Phi) \,&=\, \gamma\pi(r+s) \exp\left( -\gamma\pi(r^2-s^2) \right), \\
    \widetilde{M}_2(r,s,\Phi) \,&=\, -\gamma^2\pi (r+s)^2 \exp\left( -\gamma\pi(r^2-s^2) \right).
\end{align*}
We can also express these formulas in terms of other quantities that describe the relationship between the inner and outer radius.
There are three useful ways in which the inner and outer radius can be related to each other:

For rings with a constant factor $f={s}/{r}$, the MCFs are given by
\begin{align*}
M_0(r,s,\Phi) \,&=\, 1-\exp\left(-\gamma\pi r^{2}(1-f^{2})\right), \\
M_1(r,s,\Phi) \,&=\, \gamma\pi\, r(1+f)\, \exp\left(-\gamma\pi r^{2}(1-f^{2})\right), \\
\widetilde{M}_2(r,s,\Phi) \,&=\, -\gamma^2\pi\, r^2(1+f)^2\, \exp\left(-\gamma\pi r^{2}(1-f^{2})\right).
\end{align*}
For rings with a constant thickness, $\varepsilon=r-s$ the MCFs are given by
\begin{align*}
M_0(r,s,\Phi) \,&=\, 1-\exp\left(-\gamma\pi\varepsilon (2r-\varepsilon)\right), \\
M_1(r,s,\Phi) \,&=\, \gamma\pi\,(2r-\varepsilon)\, \exp\left(-\gamma\pi\varepsilon (2r-\varepsilon)\right), \\
\widetilde{M}_2(r,s,\Phi) \,&=\, -\gamma^2\pi\,(2r-\varepsilon)^{2}\, \exp\left(-\gamma\pi\varepsilon (2r-\varepsilon)\right).
\end{align*}
A constant area $A_{0}$ can be obtained for $r > \sqrt{{A_{0}}/{\pi}}$ by choosing the inner radius $s=\sqrt{r^{2}-{A_{0}}/{\pi}}$.
In this case, the MCFs are given by
\begin{align*}
M_0(r,s,\Phi) \,&=\, 1-\exp\left( -\gamma A_0 \right), \\
M_1(r,s,\Phi) \,&=\, \gamma\pi (r+\sqrt{r^{2}-{A_{0}}/{\pi}})\, \exp\left( -\gamma A_0 \right),\\
\widetilde{M}_2(r,s,\Phi) \,&=\, -\gamma^2\pi(r+\sqrt{r^{2}-{A_{0}}/{\pi}} )^{2}\, \exp\left( -\gamma A_0 \right).
\end{align*}
Then, the fraction of area $C:=1-\exp\left( -\gamma A_0 \right)$ covered by the rings is a constant for a Poisson point process.

\begin{figure}[t]
  \centering
  \subfigure[][]{\includegraphics[width=0.92\linewidth]{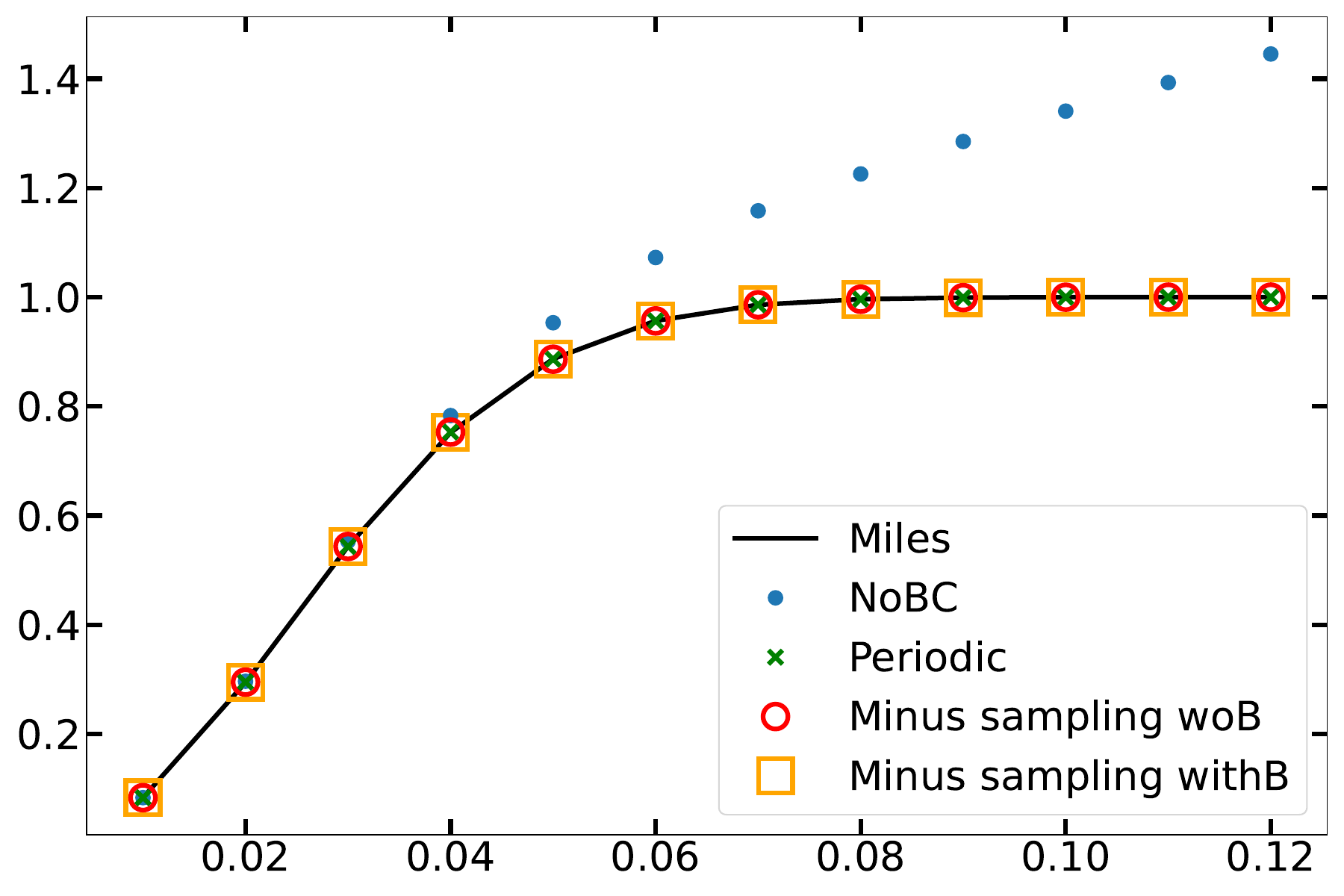}} \\
  \subfigure[][]{\includegraphics[width=0.92\linewidth]{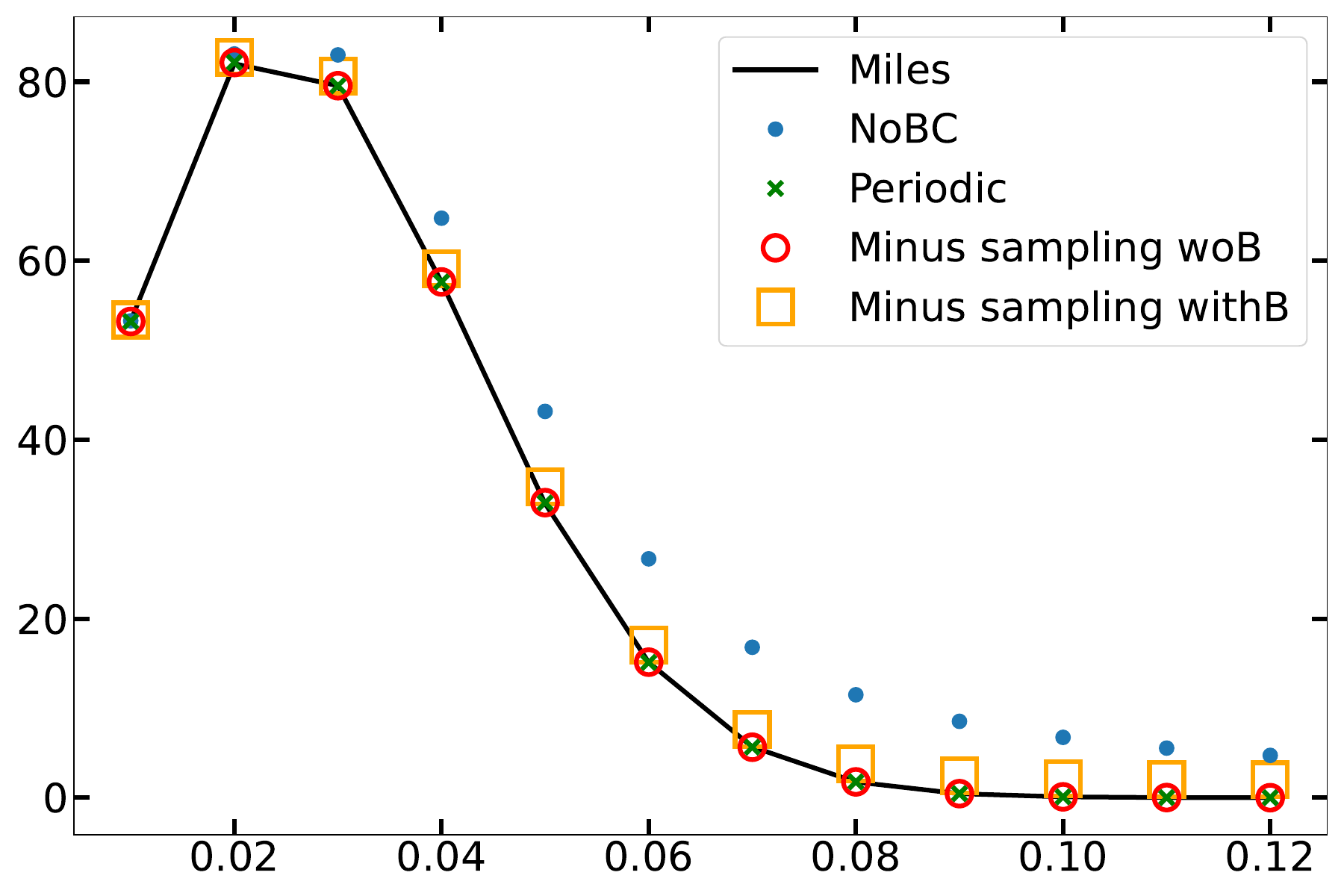}} \\
  \subfigure[][]{\includegraphics[width=0.92\linewidth]{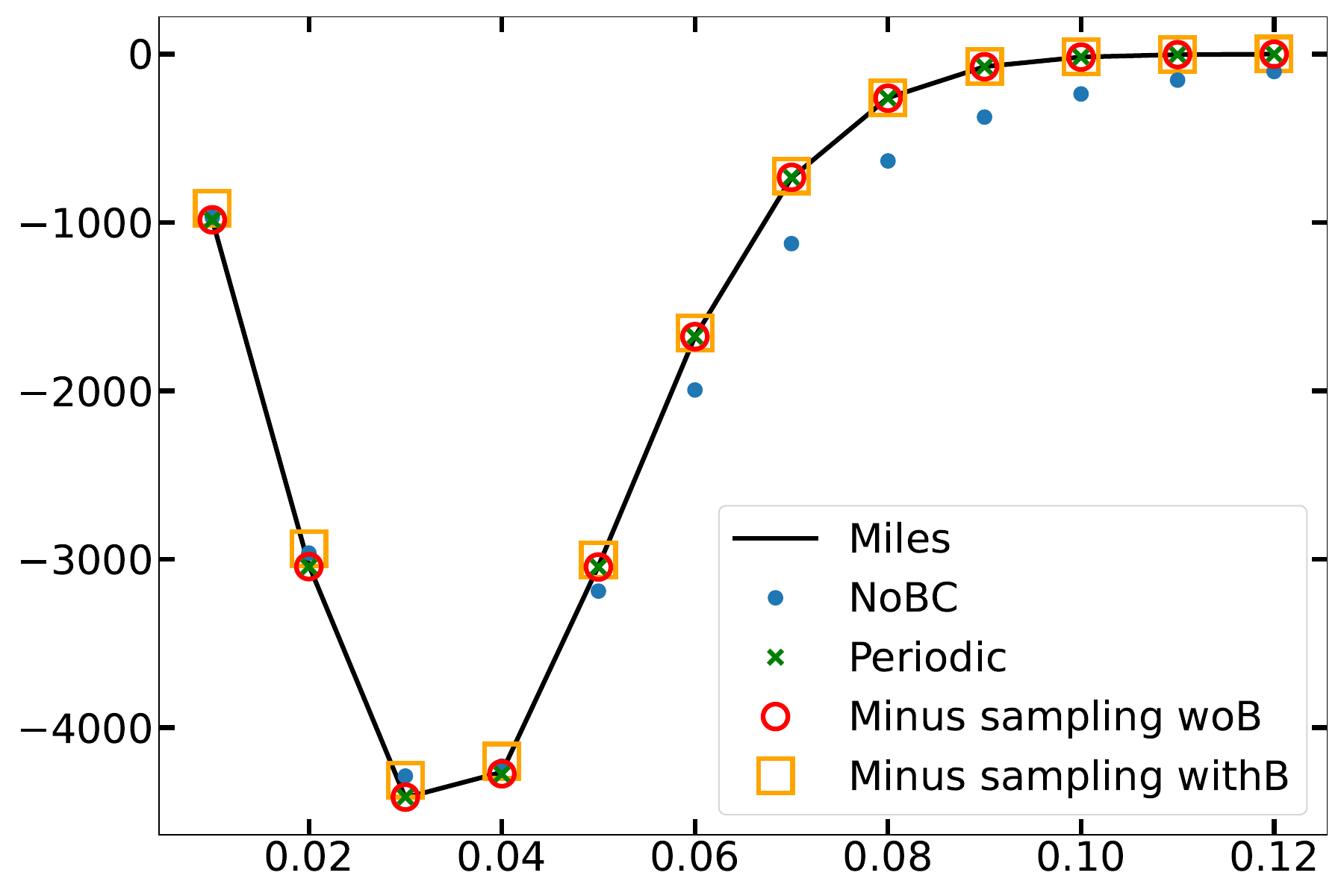}}
  \caption{Minkowski functionals (a) Area $M_0$, (b) (half the) Perimeter $M_1$ and (c) Euler characteristic for rings with constant factor $\frac{s}{r}=0.85$ as a function of the outer radius $r$ for a Poisson process of 1000 points in the unit square. For each value of $r$ we took the average of 500 independent realizations.}
  \label{fig:Poisson_MCF}
\end{figure}

Figure \ref{fig:Poisson_MCF} shows the three Minkowski correlation functions of a Poisson point process for the case of a constant factor $f$ and compares numerical to analytical results.

\section{Robust arbitrary multipoint information}
\label{Sec:multipoint}

\subsection{Relation between Euler characteristic and pair correlation function}\label{Sec:EulerPairCor}

The intersection of two, three, or in general $n$ rings, contains information about the correlation between two, three, or $n$ points, respectively.
Therefore, their Minkowski functionals and the MCFs contain information about correlations of arbitrary order.
This is also evident from formula~\eqref{eq:Formula_MCF}.

Expressing the MCFs by multipoint correlation functions reveals the importance of the choice of thickness $\varepsilon=r-s$ of the rings.
Heuristically, it allows to control the degree of higher- or lower-order statistics that is contained in the MCFs. 
For not too small values of $\varepsilon$, the Minkowski functionals contain the intersection of three or more rings and thus higher-order statistics.
If $\varepsilon$ is chosen so small that the intersection of three or more rings does not occur, the MCFs reduce to a second-order statistic and contain mainly the information of the 2-point correlation function $\rho_2$.
The latter can be derived directly from the Euler characteristic.

For $\varepsilon\rightarrow0$ (more precisely $s\uparrow r$), all terms $n>2$ in Eq.~\eqref{eq:Formula_MCF} vanish because there are no intersections of that order (and the Minkowski functionals of the empty set are zero).
For the Euler characteristic $M_2$, also the first term vanishes, because the Euler characteristic vanishes for a single ring.
Therefore, only the second term, which includes the pair-correlation function, does not vanish \cite{Mecke_workingpaper_2008}:
\begin{align*}
    \lim_{s\uparrow r}\, M_2(r,s,\Phi) \,&=\, \sum_{n=1}^\infty \frac{(-1)^{n+1}}{n!} \int_{(\R^2)^{n-1}} \rho_n(\0,x_2,\dots,x_{n}) \\
    &\hspace*{-6pt} \cdot \lim_{s\uparrow r}\, M_2\left(R(\0)\cap \cap_{i=2}^{n} R(x_i) \right) \;\d{(x_2,\dots,x_{n})} \\
    &\hspace*{-18pt}=\, -\frac{1}{2} \int_{\R^2} \rho_2(\0,x) \lim_{s\uparrow r}\, M_2\left(R(\0)\cap R(x) \right) \;\d{x}.
\end{align*}
The interchange of limit and integral can be justified using the bound~\eqref{Euler:upperBound}.
For 0<$\Vert x\Vert<2r$, the Euler characteristic of the intersection $R(\0)\cap R(x)$ in the limit $s\uparrow r$ equals 2 and therefore
\begin{align*}
    \lim_{s\uparrow r}\, M_2(r,s,\Phi) \,=\, - \int_{B(\0,2r)} \rho_2(\0,x) \;\d{x}.
\end{align*}
For a statistically isotropic and homogeneous point process, the 2-point correlation function $\rho_{2}(x,y)$ depends only on the distance $\Vert x-y\Vert$.
Using $\rho_2(\0,x) = \gamma^2 g^{(2)}(\Vert x\Vert)$ for stationary isotropic processes, we obtain
\begin{align*}
    \lim_{s\uparrow r}\, M_2(r,s,\Phi) \,=\, -2\pi\gamma^2 \int_{0}^{2r} g^{(2)}(t)\, t \;\d{t}.
\end{align*}
In particular, the derivative of the limiting Euler characteristic with respect to $r$ is, up to a known $r$-dependent prefactor, proportional to the pair-correlation function $g^{(2)}$ evaluated at separation $2r$.

This is a proof of principle that the $n$-point correlation functions can be extracted from the MCFs.
However, we want to emphasize that we do not suggest a new approach to calculate these standard correlation functions.
Rather, we propose the MCFs as a novel approach that as additive valuations  contains $n$-point information in a statistically robust way. 
For $\varepsilon >0$, the MCFs do contain higher order correlations and can distinguish point processes with the same pair-correlation function and thus with the same structure factor.

In the next subsection we show how for $\varepsilon >0$, the MCFs distinguish two processes with the same second order properties.

\subsection{Baddeley--Silverman process}
\label{sec:baddeley-silverman}

For a finite thickness of the rings, the morphological functions contain information beyond second-order statistics.
To illustrate this property, we consider the Baddeley--Silverman process, a classical example of a point process that has the same pair correlation function, and consequently the same structure factor, as a homogeneous Poisson process, while differing in its higher-order spatial structure~\cite{mecke_morphological_2005}.

The process can be constructed in a particularly simple way.
We partition the plane into unit squares and independently assign to each square a random number $N$ of points according to
\[
    \mathbb{P}(N=0)=\frac{1}{10}, \quad
    \mathbb{P}(N=1)=\frac{8}{9}, \quad
    \mathbb{P}(N=10)=\frac{1}{90}.
\]
Conditional on $N$, the points are placed independently and uniformly inside the corresponding square.
Since
\[
    \mathbb{E}[N]=1
    \qquad\text{and}\qquad
    \operatorname{Var}(N)=1,
\]
the first two moments of the number of points in each unit square agree with those of a Poisson random variable with mean one. 
The resulting process has intensity $\gamma=1$ and the same pair correlation function as a Poisson point process.
Its local structure, however, is clearly different: most cells contain a single point, some cells are empty, and rare cells contain ten points, producing small, dense clusters.
To obtain a stationary point process rather than one tied to a fixed location of the underlying grid, the cell structure is shifted by a uniformly distributed random vector within one unit square.

\begin{figure}[t]
    \centering
    \includegraphics[width=\linewidth]{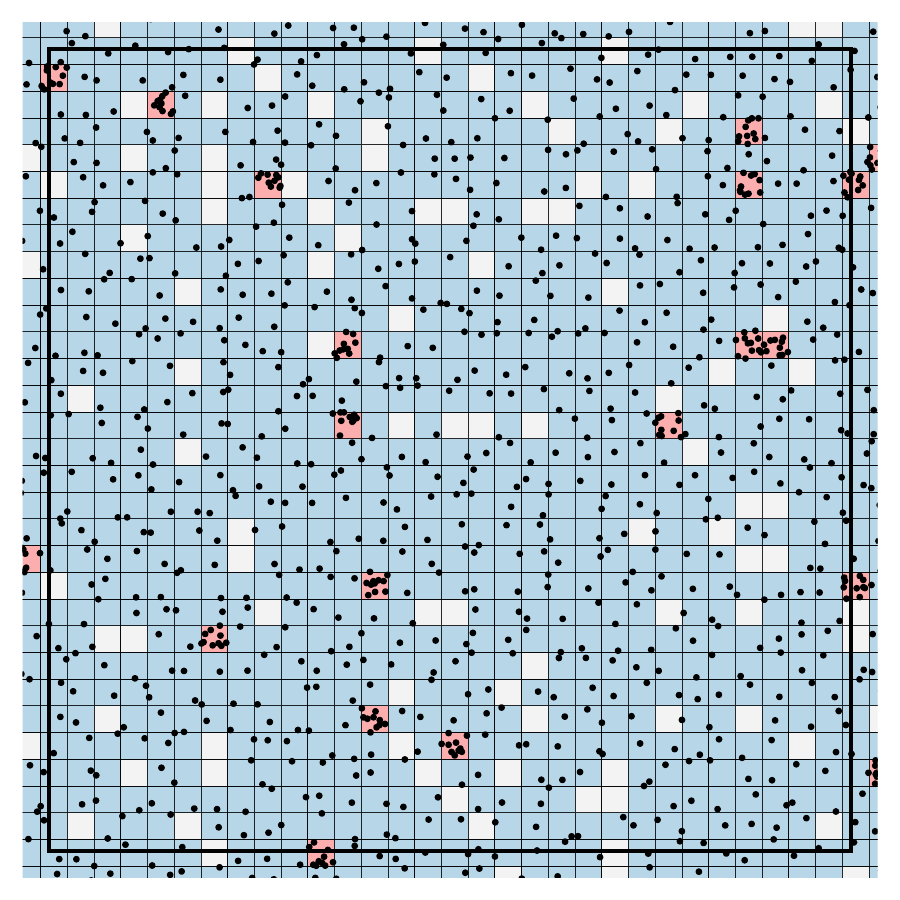}
    \caption{Realization of the Baddeley--Silverman process in the observation window $W=[0,30]^2$. The colors of the underlying unit cells indicate whether a cell contains zero, one, or ten points.}
    \label{fig:baddeley-pattern}
\end{figure}

\begin{figure}[t]
    \centering
    \includegraphics[width=\linewidth]{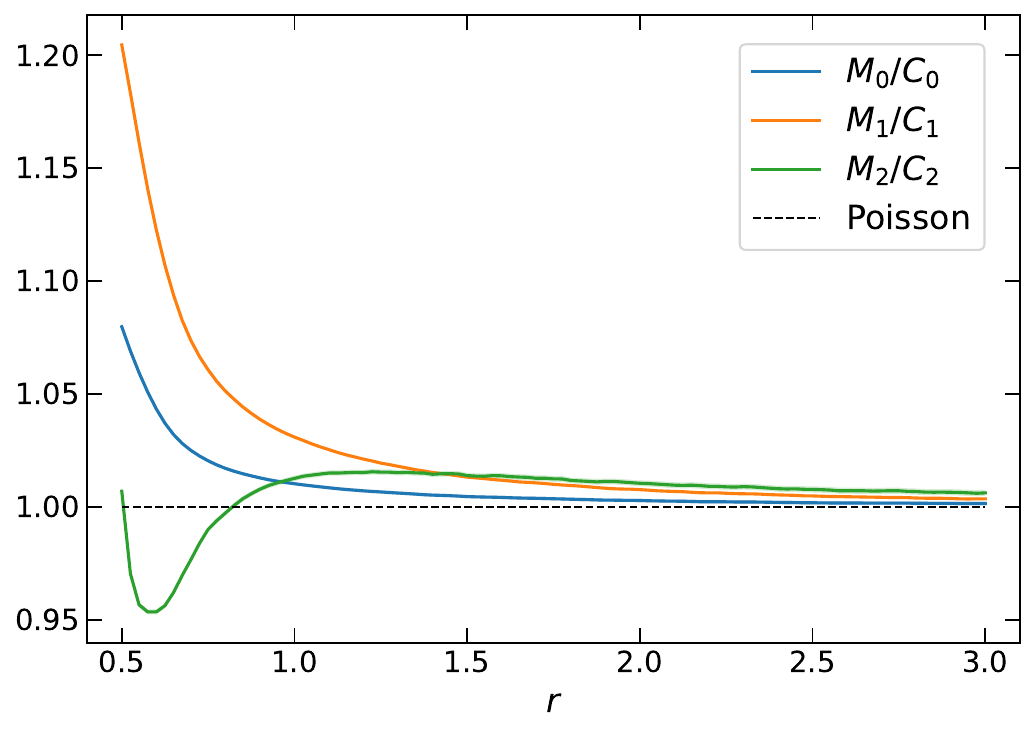}
    \caption{Normalized MCFs for the Baddeley--Silverman process as functions of the outer radius $r$, with the inner radius $s$ chosen according to \eqref{s_normalization}.
    The curves are averages over $500$ independent realizations.
    The dashed line denotes the corresponding Poisson reference.}
    \label{fig:baddeley-minkowski}
\end{figure}

Figure~\ref{fig:baddeley-pattern} illustrates one realization in the observation window $W=[0,30]^2$.
The underlying cells are included in the figure to visualize the construction.
Figure~\ref{fig:baddeley-minkowski} shows the normalized MCFs as a function of the outer radius $r$, averaged over $500$ independent realizations of the Baddeley--Silverman process.
For each $r$, the inner radius $s=s(r)$ is chosen such that the area of the ring $R$ is independent of $r$,
\begin{align} \label{s_normalization}
    M_0(R) \,=\, \pi\bigl(r^2-s^2\bigr) \,=\, \log 2.
\end{align}
For a homogeneous Poisson process of intensity 1, this choice corresponds to an expected covered area fraction of $1/2$.
The MCFs are computed using minus sampling without including contributions from the boundary of the observation window.
For each radius $r$, the point pattern is therefore sampled in $[-r,30+r]^2$, such that the corresponding reduced observation window is exactly $[0,30]^2$.
We normalize the three MCFs by
\begin{align*}
    C_0(r) &= \frac{1}{2},\\
    C_1(r) &= \frac{\pi}{2}\gamma(r+s),\\
    C_2(r) &= -\frac{\pi}{2}\gamma^{\,2}
              (r+s)^2.
\end{align*}
With this normalization, the Poisson reference equals one for all three quantities $M_0/C_0$, $M_1/C_1$, and $M_2/C_2$.

The morphological functions clearly distinguish the Baddeley--Silverman process from its Poisson counterpart despite their identical pair correlation functions.
The deviations are strongest at small ring radii.
However, deviations from the Poisson reference remain visible even at larger radii, confirming that the MCFs also capture spatial correlations over longer ranges.
In this context, recall that the cells of the Baddeley--Silverman process have side length 1, so that a ring of radius 2 already spans approximately four cells.
In particular, the functions $M_1$ and $M_2$ of the Baddeley--Silverman process remain distinguishable from those of the Poisson process even at relatively large radii.
The Baddeley--Silverman example thus provides a direct demonstration that the MCFs contain spatial information beyond pair correlations. 
Such higher-order information is particularly relevant for
the analysis of physical point patterns, whose characteristic spatial
structure is often reflected in higher-order statistics
\cite{Mecke_Integralgeometrie_1994,MeckeOverview2000,mecke_morphological_2005}.

\begin{figure}[t]
    \centering
    \includegraphics[width=\linewidth]{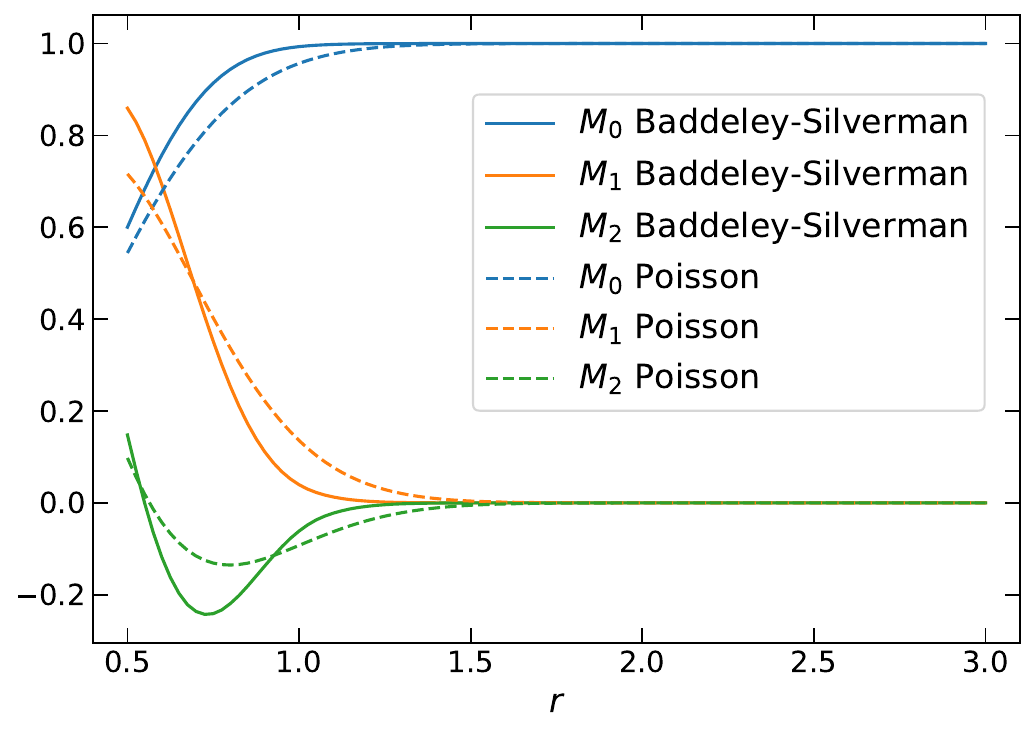}
    \caption{MCFs for the Baddeley--Silverman process as functions of the outer radius $r$ with inner radius $s=0$.
    The curves are averages over $500$ independent realizations.
    The dashed lines denotes the corresponding Poisson reference.}
    \label{fig:baddeley-minkowski_s=0}
\end{figure}

Finally, we compare the range of spatial scales over which the MCFs distinguish the Baddeley--Silverman process from the Poisson process with that of the previously established correlation functions introduced in~\cite{mecke_robust_1994}, where the points are decorated with balls instead of rings, corresponding to the special case $s=0$ in our setting \cite{mecke_morphological_2005}.
To this end, we additionally computed the MCFs of the Baddeley--Silverman process for $s=0$ and compared them with the corresponding Poisson reference.
Since balls rapidly cover almost the entire space as their radius increases, both $M_1$ and $M_2$ become very small.
We therefore refrain from normalizing the MCFs by the Poisson reference values $C_j(r)$ in this comparison.
As shown in Fig.~\ref{fig:baddeley-minkowski_s=0}, for $s=0$ the MCFs of the Baddeley--Silverman and Poisson processes become practically indistinguishable for radii larger than approximately $1.7$.
In contrast, the ring-based MCFs with $s>0$ retain a clear distinction between the two processes over substantially larger radii; see Fig.~\ref{fig:baddeley-minkowski}.

\section{Applications to scientific data}
\label{sec:applications}

In the previous sections, we investigated the morphological characteristics of point process models for which the underlying spatial structure is known.
We now apply the morphological analysis to three point patterns obtained from scientific data.
The examples span very different spatial scales, ranging from a microscopic particle system to the locations of trees and an astronomical point pattern.
This illustrates that the MCFs can be applied to point patterns arising in very different scientific contexts.

For each data set, we estimate the point intensity by
\begin{align*}
    \widehat{\gamma} = \frac{N}{|W|},
\end{align*}
where $N$ is the number of observed points and $W$ is the observation window.
To make the analysis comparable across the different spatial scales and intensities, distances are expressed in terms of the dimensionless radius $\sqrt{\widehat{\gamma}}\,r$.
As in Sec.~\ref{sec:baddeley-silverman} we choose the inner radius $s=s(r)$ of each ring such that its area is independent of $r$ and the expected covered area fraction is $1/2$ for a homogeneous Poisson process of intensity $\widehat{\gamma}$.

\begin{figure*}[t]
    \centering
    \includegraphics[width=0.96\textwidth]{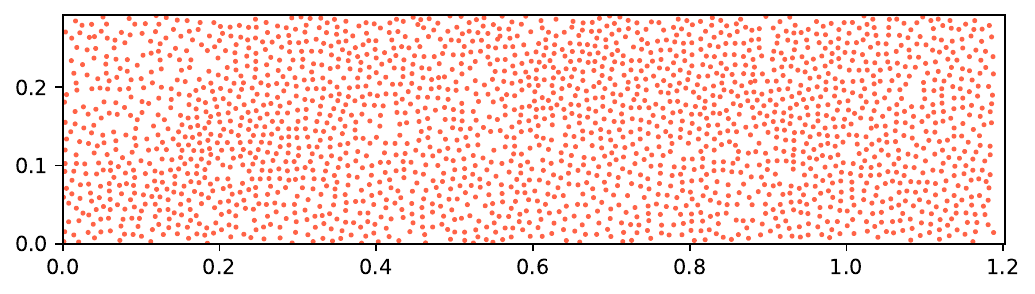}
    \caption{$2137$ Quincke rollers in an experimental snapshot}
    \label{fig:quincke}
\end{figure*}

Since the data are observed in bounded and non-periodic windows, we use the minus boundary condition introduced in Sec.~\ref{Sec:BoundaryConditions} without taking contributions of the boundary of the observation window into account.
For a rectangular observation window
\begin{align*}
    W=[x_{\min},x_{\max}]\times[y_{\min},y_{\max}],
\end{align*}
the effective observation window at outer ring radius $r$ is given by
\begin{align*}
    W_{\ominus r}
    =
    [x_{\min}+r,x_{\max}-r]
    \times
    [y_{\min}+r,y_{\max}-r].
\end{align*}
As before, the normalization factors are chosen such that all three corresponding Poisson
reference values are equal to one:
\begin{align*}
    C_0(r) &= \frac{1}{2}|W_{\ominus r}|,\\
    C_1(r) &= \frac{\pi}{2}\widehat{\gamma}(r+s)|W_{\ominus r}|,\\
    C_2(r) &= -\frac{\pi}{2}\widehat{\gamma}^{\,2}
              (r+s)^2|W_{\ominus r}|.
\end{align*}
Their deviations from unity therefore provide a direct representation of morphological differences between the observed point pattern and a homogeneous Poisson
process of the same intensity.

\subsection{Quincke particle system}
\label{sec:quincke}

Our first application considers an experimental point pattern of $2137$ Quincke rollers in a microfluidic channel.
Quincke rollers are micrometer-sized colloidal particles that self-propel along a surface due to an electrohydrodynamic instability and provide an experimental realization of an active polar particle system \cite{Geyer2018,Supekar2023}.
The particles have a diameter of $4.8\,\mu\mathrm{m}$ and can reach velocities of the order of millimeters per second.
The particle positions analyzed here are taken from the experimental data used by Supekar et al.~\cite{Supekar2023} to infer hydrodynamic equations for the collective dynamics of Quincke rollers.
We consider one snapshot containing $2137$ particles in an observation window of approximately $1.203\,\mathrm{mm}\times0.292\,\mathrm{mm}$.
Figure~\ref{fig:quincke} shows the corresponding particle positions.

\begin{figure}[t]
    \centering
    \includegraphics[width=\linewidth]{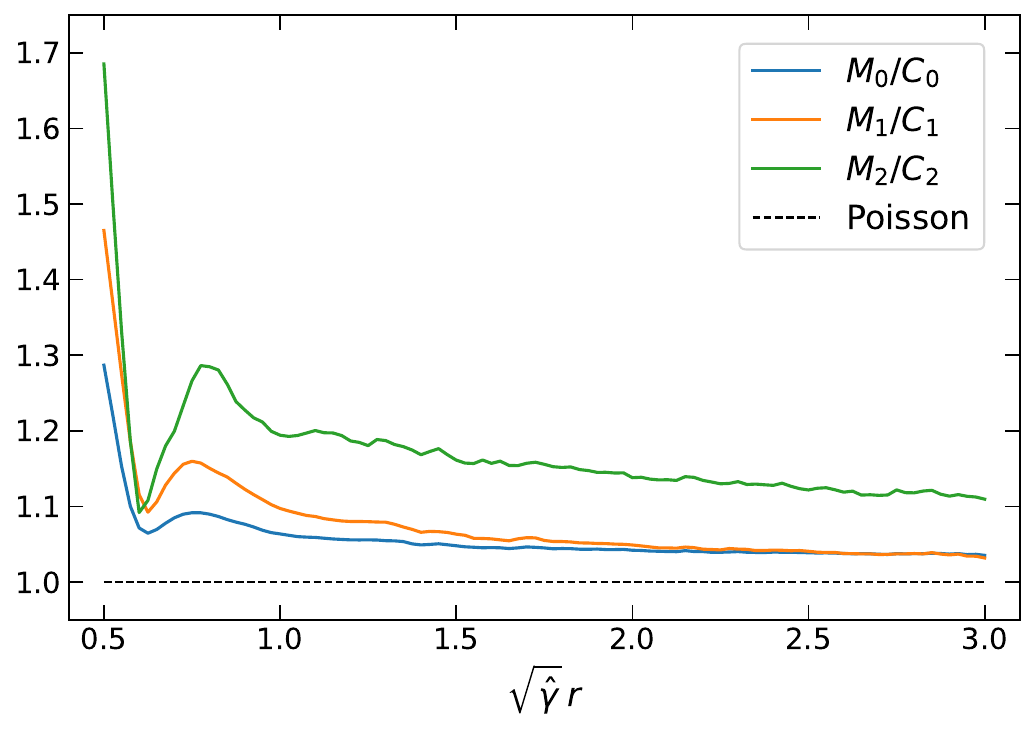}
    \caption{Normalized MCFs for the Quincke particle pattern. The dashed line denotes the corresponding
    Poisson reference.}
    \label{fig:quincke-minkowski}
\end{figure}

The corresponding normalized MCFs are shown in Fig.~\ref{fig:quincke-minkowski}.
All three morphological functions are above the Poisson reference over the considered range of
scaled radii.
The deviation is already visible for $M_0/C_0$ and $M_1/C_1$, but is considerably more pronounced for $M_2/C_2$.

The deviations from the Poisson reference are consistent with the comparatively regular spatial arrangement of the particles relative to a homogeneous Poisson pattern.
In particular, the values of $M_0/C_0$ and $M_1/C_1$ above unity can be interpreted geometrically in terms of reduced overlap between neighboring rings compared with the Poisson reference.
The substantially stronger response of $M_2/C_2$ highlights the sensitivity of the Euler characteristic to changes in the topology of the union, in particular to the formation and merging of connected components and holes.
The MCFs therefore reveal structural differences from a homogeneous Poisson pattern across all three Minkowski functionals, with the most pronounced response occurring in the topological descriptor $M_2$.

\subsection{Longleaf pine trees}
\label{sec:longleaf}

\begin{figure}[t]
    \centering
    \includegraphics[width=0.48\textwidth]{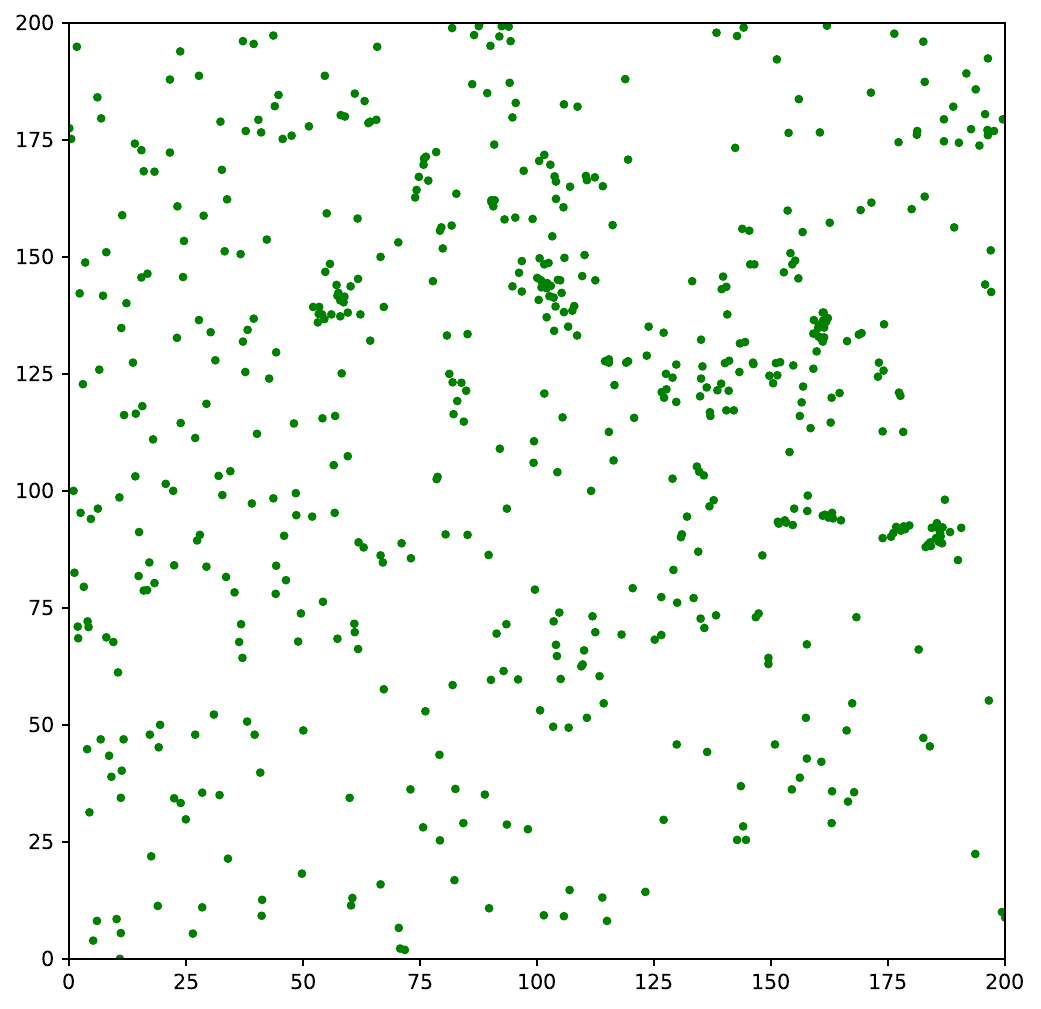}
    \caption{$584$ longleaf pine trees in Georgia, USA}
    \label{fig:longleaf}
\end{figure}

Our second application considers the locations of longleaf pine trees (\textit{Pinus palustris}) in the Wade Tract, an old-growth forest in Thomas County, Georgia, USA~\cite{Platt1988}.
The data set analyzed here contains $584$ tree locations in a $200\,\mathrm{m}\times200\,\mathrm{m}$ observation window.
The point pattern has previously been studied using Minkowski functionals of discs in~\cite{mecke_morphological_2005,Mecke_workingpaper_2008,Bittner_diploma_2008}
Figure~\ref{fig:longleaf} shows the observed tree locations.
The pattern is spatially heterogeneous, with groups of closely spaced trees separated by more sparsely populated regions.

\begin{figure}[t]
    \centering
    \includegraphics[width=\linewidth]{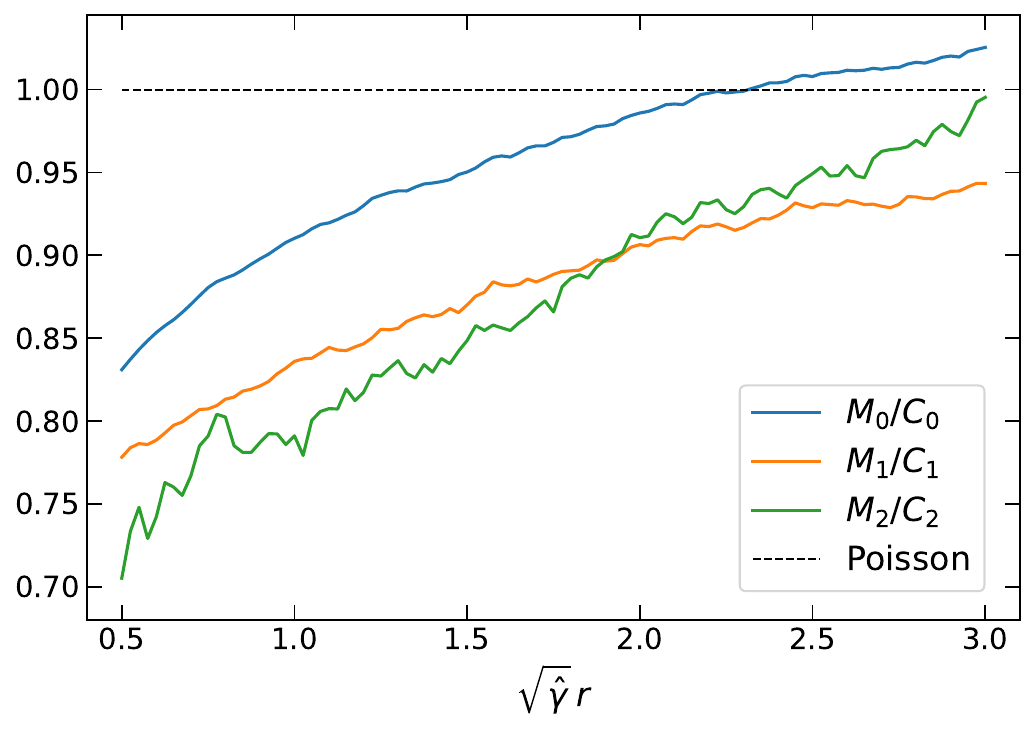}
    \caption{Normalized MCFs for the longleaf pine pattern. The dashed line denotes the corresponding
    Poisson reference.}
    \label{fig:longleaf-minkowski}
\end{figure}

The corresponding normalized MCFs are shown in Fig.~\ref{fig:longleaf-minkowski}.
At small scaled radii, all three morphological functions lie substantially below the Poisson
reference.
As the ring radius increases, the functions gradually approach
unity.

The deviations from the Poisson reference are consistent with the spatial heterogeneity visible in the point pattern.
In regions with locally high tree density, rings centered at neighboring trees overlap more strongly than in more sparsely populated regions.
Such increased overlap provides a geometric explanation for the reduced covered area and total boundary length of the union and hence for values of $M_0/C_0$ and $M_1/C_1$ below unity at small and intermediate radii.
All three MCFs reveal clear morphological differences from a homogeneous Poisson pattern.

\subsection{Hubble Ultra Deep Field}
\label{sec:hubble}

\begin{figure}[t]
    \centering
    \includegraphics[width=0.48\textwidth]{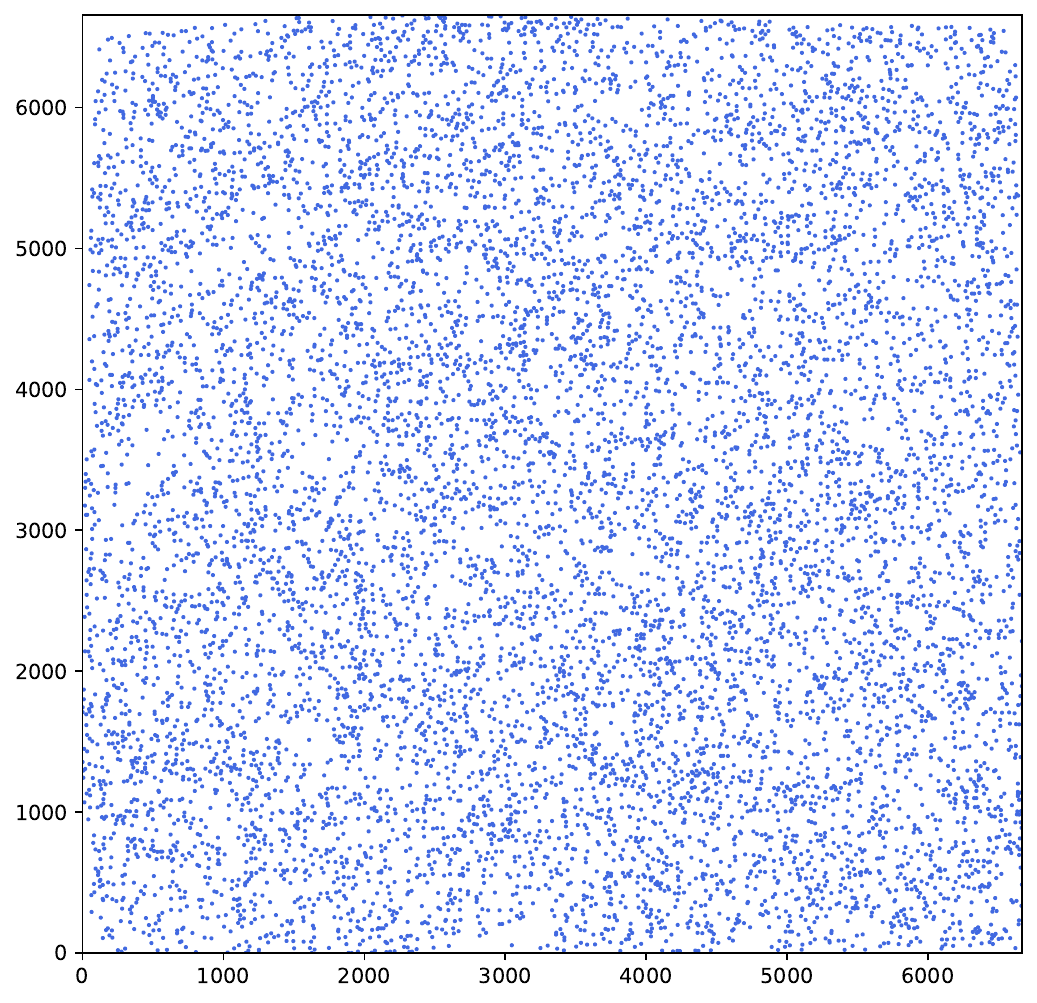}
    \caption{Hubble Ultra Deep Field point pattern after rotation and translation}
    \label{fig:hubble}
\end{figure}

Our third application considers an astronomical point pattern from the
Hubble Ultra Deep Field (HUDF).
The HUDF is a deep-field imaging survey obtained with the Wide Field Camera of the Advanced Camera for Surveys aboard the Hubble Space Telescope.
We use the source catalog derived from the F775W ($i$-band) image~\cite{STScI2004HUDF}, which contains $10179$ detected astronomical sources whose positions are given in pixel
coordinates.
The corresponding drizzled image has dimensions $10500\times10500$ pixels with a pixel scale of $0.03''$ per pixel.
The original catalog has an approximately square-shaped footprint that is rotated with respect to the coordinate axes.
To obtain a rectangular observation window suitable for the minus-sampling procedure, we aligned the point pattern with its minimum-area bounding rectangle, corresponding to a rotation of approximately $42.18^\circ$, and translated the resulting rectangle to the origin.
This yields the observation window
\begin{align*}
    W=[0,6666.93]\times[0,6657.41]
\end{align*}
in pixel coordinates.
The resulting point pattern, which is used for the subsequent analysis, is shown in Fig.~\ref{fig:hubble}.
Since no separate observation mask is available, this rectangle should be regarded as a data-driven approximation of the effective observational footprint.
The applied rotation and translation are rigid motions and therefore preserve distances and the morphological structure of the point pattern.

\begin{figure}[t]
    \centering
    \includegraphics[width=\linewidth]{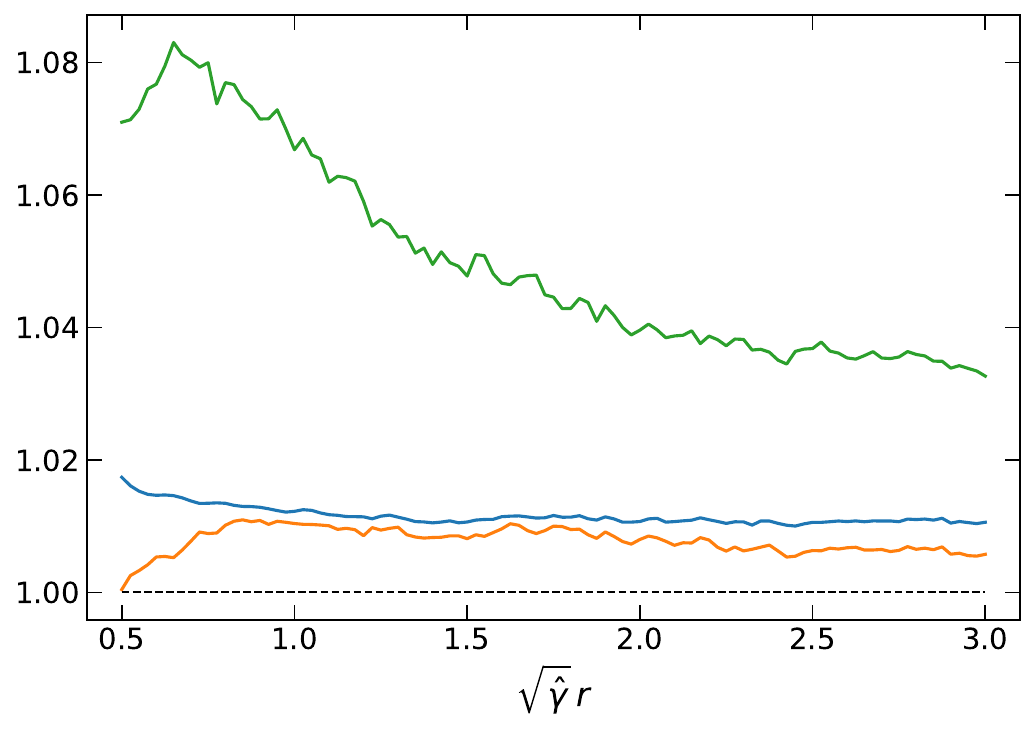}
    \caption{Normalized MCFs for the Hubble Ultra Deep Field point pattern. The dashed line denotes the corresponding Poisson reference.}
    \label{fig:hubble-minkowski}
\end{figure}

The normalized MCFs for the transformed point pattern are shown in Fig.~\ref{fig:hubble-minkowski}.
The intensity $\widehat{\gamma}$ is computed from the transformed catalog within the estimated rectangular window.

All three normalized MCFs show systematic deviations from the Poisson reference throughout the entire range of considered radii.
While the deviations of $M_0/C_0$ and $M_1/C_1$ are comparatively small, $M_2/C_2$ exhibits substantially stronger deviations.
Particularly at larger radii, however, these deviations may partly result from the approximate observation window and the corresponding intensity estimate rather than from spatial correlations.
The HUDF data thus illustrate the potential of MCFs for detecting morphological differences from a Poisson reference, while highlighting the importance of accurately specifying the observational footprint.

\section{Conclusions and outlook}
\label{sec:conclusions}

We introduced  Minkowski correlation functions (MCFs) as ring-based geometric descriptors for the characterization of disordered point patterns and their spatial correlations.
By associating with each point a ring of outer radius $r$ and inner radius $s$ and characterizing the resulting union of rings through its Minkowski functionals, the MCFs provide a complementary approach to conventional $n$-point correlation functions.
In particular, they combine a multiscale geometric description with sensitivity to higher-order correlations that are not captured by the pair correlation function or the structure factor.
The underlying Minkowski functionals have proved useful as geometric and morphological descriptors in a wide range of applications~\cite{MeckeWagner1991,Mecke1998,MeckeStoyan2000,Jacobs_dewetting_1998,herminghaus_spinodal_1998,SchroederTurketal2011AdvMater,Klatt2016}.
The present construction extends the earlier disc-based approach~\cite{mecke_robust_1994,mecke_morphological_2005} by assigning rings of finite thickness to the points.
By avoiding the rapid covering of the observation window by overlapping discs, the use of rings substantially extends the range of spatial scales accessible to the morphological analysis.

On the theoretical side, we established the existence of specific MCFs in the thermodynamic limit and showed that the resulting quantities are independent of the chosen observation-window sequence.
We derived a series representation in terms of product densities which explicitly demonstrates that the MCFs generally incorporate correlations of arbitrary order.
Using results from positive-reach geometry, we further showed that the required Minkowski functionals are well defined for the finite unions and intersections of rings arising in the construction.
In the thin-ring limit $s\uparrow r$, we related the MCF based on the Euler characteristic explicitly to the pair correlation function $g^{(2)}$, thereby showing how conventional second-order information is embedded in the morphological description.
For the homogeneous Poisson process, we derived closed-form expressions for $M_0$ and $M_1$ and used a Miles-type expression for $M_2$ as an analytical reference.

For the numerical evaluation of the MCFs, we developed an exact and efficient algorithm based on local arc contributions along the boundary of the union of rings.
The algorithm evaluates the three Minkowski functionals $M_0$, $M_1$, and $M_2$ and supports boundary conditions for both periodic and non-periodic point patterns.
The additivity and motion invariance inherited from the Minkowski functionals, together with the local nature of the numerical evaluation, make the resulting descriptors particularly convenient for the geometric analysis of point patterns.
Moreover, representing the MCFs as functions of the ring radii provides an intuitive way to compare different geometrical and topological properties across spatial scales.

The numerical examples demonstrate the complementary information contained in the different MCFs and, in particular, their sensitivity to higher-order spatial structure.
The Baddeley--Silverman process provides a direct demonstration of this property, since it has the same pair correlation function and structure factor as a homogeneous Poisson process while exhibiting systematic deviations from the corresponding Poisson reference in its MCFs.
The MCFs can therefore distinguish point processes that are indistinguishable by these conventional second-order descriptors without requiring a parametric model for their higher-order correlations.
Furthermore, we observed that, for $s=0$, corresponding to the use of balls, the MCFs of the Baddeley--Silverman process become practically indistinguishable from those of the Poisson process beyond a certain radius, whereas the use of genuine rings allows the two processes to be distinguished over a larger range of radii.

Our applications to experimental and observational data further demonstrate the applicability of the approach to point patterns arising on vastly different physical scales.
For active Quincke rollers, longleaf pine trees, and astronomical sources in the Hubble Ultra Deep Field, the MCFs reveal systematic morphological deviations from a homogeneous Poisson process of the same intensity.
The three Minkowski functionals respond to different aspects of the geometry and topology of the point patterns, so that covered area, boundary length, and Euler characteristic provide complementary information about their spatial structure.

Several extensions of the present framework provide promising directions for future work.
One natural direction is the extension to higher-dimensional point patterns and to more general grain shapes, for which the corresponding intrinsic volumes provide higher-dimensional analogues of the planar Minkowski functionals considered here.
Another natural generalization concerns point patterns on curved spaces.
In particular, Minkowski functionals can also be defined on the sphere, opening the possibility of constructing analogous correlation functions for spherical or directional point patterns.

A further extension concerns anisotropic point patterns, whose directional structure cannot be fully characterized by scalar Minkowski functionals.
Minkowski tensors provide tensor-valued generalizations of the Minkowski functionals and can systematically quantify directional structure and anisotropy~\cite{SchroederTurketal2011AdvMater}.
Incorporating Minkowski tensors or scalar anisotropy indices derived from them into the present correlation framework could therefore provide multiscale descriptors of anisotropic spatial correlations.

Another important direction is the development of statistical inference based on MCFs.
Confidence bands, resampling procedures, and formal hypothesis tests against Poisson or fitted reference models could supplement morphological comparisons by quantitative measures of statistical uncertainty.
The choice of the inner and outer ring radii also offers scope for further investigation, including systematic strategies for selecting the ring thickness or area and the investigated range of spatial scales according to the resolution and characteristic structures of a given data set.

Taken together, the ring-based MCFs provide a unified geometric framework for probing spatial correlations across length scales and beyond conventional second-order descriptors.
Their combination of higher-order sensitivity, complementary geometrical and topological information, and efficient computation makes them a versatile addition to established methods for the analysis of complex point patterns and provides several directions for further theoretical, statistical, and computational development.

\begin{acknowledgments}
We thank Sebastian Bittner and Uwe Brodatzki for their valuable contributions during the early stages of this project, which provided an important foundation for the present work \cite{Brodatzki_diss_2003,Bittner_diploma_2008}.
This work was financially supported by the Deutsche Forschungsgemeinschaft (DFG, German Research Foundation) through the SPP 2265, under grant number ME 1361/16-1 and KL 3391/2-2.
This work was also supported by the Initiative and Networking Fund of the Helmholtz Association under the call Helmholtz Young Investigator Groups (VH-NG-19-34, DataMat).
The authors gratefully acknowledge the scientific support and HPC resources provided by the Erlangen National High Performance Computing Center (NHR@FAU) of the Friedrich-Alexander-Universität Erlangen-Nürnberg (FAU).
This research has made use of the VizieR catalogue access tool, CDS,
Strasbourg, France \cite{10.26093/cds/vizier}.
The original description
of the VizieR service was published in \cite{vizier2000}.
\end{acknowledgments}

\appendix
\onecolumngrid

\section*{Appendix}

The appendix collects the technical arguments underlying the theoretical results of the paper.
Section~\ref{App:Technic} establishes uniform bounds for Minkowski functionals of finite intersections of rings and uses them to justify the applications of Fubini's theorem and the dominated convergence theorem in the series representation of the MCFs.
Section~\ref{App:proof} contains the proof of Theorem~\ref{Th:positiveReachIntersections} on the positive reach of finite intersections of rings.
Finally, Section~\ref{App:MilesProof} contains the proof of Lemma~\ref{Lem:Miles} for the Poisson reference.

Throughout the appendix, we fix radii $0<s<r$ and denote by $R(x)$ the closed ring centered at $x\in\mathbb{R}^d$ with inner radius $s$ and outer radius $r$.
We write $B(x,t)$ for the closed ball centered at $x\in\mathbb{R}^d$ with radius $t>0$, and $\kappa_d$ for the $d$-dimensional volume of the unit ball, such that $|B(x,t)|=\kappa_d t^d$.
Whenever the Euler characteristic $M_2$ is considered in the arguments below, we restrict to $d=2$.
The $n$-th product density of the stationary point process is denoted by $\rho_n$, and we assume throughout that the uniform bound~\eqref{Korfkt:uniformlybounded} holds.

\section{Technical Assumptions and Justifications}\label{App:Technic}

This section provides a verification of the mathematical assumptions needed in Section~\ref{Sec:DefinitionMCF} for the application of Fubini’s theorem and the dominated convergence theorem.
What we prove holds for the volume $M_0$ and the (half) surface area $M_1$ in any dimension.
For the Euler characteristic $M_2$, we restrict to $d=2$, where the required bound can be obtained from the Gauss--Bonnet theorem.
To this end, we require suitable upper bounds for the absolute values of the Minkowski functionals of finite intersections of rings $R(x_1),\dots,R(x_n)$.

The volume of the intersection can be trivially bounded by the volume of one of the rings.
Since the boundary of $\cap_{i=1}^n R(x_i)$ consists of subsets of the boundaries of the individual rings $R(x_i)$, we obtain the upper bounds
\begin{align} \label{VolumeSurface:upperBound}
    M_{j}\big( \cap_{i=1}^n R(x_i) \big) \,\leq\, \sum_{i=1}^n M_{j}(R(x_i)) \,=\, nM_{j}(R(\0)), \qquad j\in\{0,1\},
\end{align}
where the factor $n$ can be omitted for $j=0$.
In the case $d=2$ the boundary of $\cap_{i=1}^n R(x_i)$ is a piecewise smooth curve and we can apply the Gauss--Bonnet theorem.
Each nonsmooth boundary point is an intersection point of the boundaries of at least two rings.
Since the boundaries of two rings can have at most 8 intersection points, the number of nonsmooth boundary points can be bounded from above by $8\binom{n}{2}$.
All in all we get the upper bound
\begin{align} \label{Euler:upperBound}
    \big| M_2\big( \cap_{i=1}^n R(x_i) \big) \big| \,\leq\, \frac{1}{\pi s} M_{1}\big( \cap_{i=1}^n R(x_i) \big) + 8\binom{n}{2} \,\leq\, \frac{n}{\pi s} M_{1}(R(\0)) + 8\binom{n}{2}.
\end{align}
For almost all $(x_1,\dots,x_n)\in(\R^d)^n$, the bounds \eqref{VolumeSurface:upperBound} and \eqref{Euler:upperBound} hold, which suffices for the following lemma.

\begin{Lem} \label{Lem:FUBDOM}
Let $n\in\N$, $\rho_n$ be the $n$-th correlation function of a stationary point process on $\R^d$ satisfying \eqref{Korfkt:uniformlybounded} and $W\subset\R^d$ a bounded, measurable set with $|W|>0$.
Furthermore define
\begin{align*}
    a_n \,:=\, n  C^n D^{n-1} M_0(R(\0)), \qquad b_n \,:=\, n  C^n D^{n-1} M_1(R(\0)), \qquad c_n \,:=\, n C^n D^{n-1} \left(\frac{1}{\pi s}M_{1}(R(\0)) + 4(n-1)\right),
\end{align*}
where $C$ is the constant from \eqref{Korfkt:uniformlybounded} and $D:=|B(\0,2r)|=(2r)^d\kappa_d$.
Then
\begin{align*}
    \frac{1}{|W|} \int_{W^n} M_0\big(\cap_{i=1}^n R(x_i)\big) \rho_n(x_1,\dots,x_n) \;\d{(x_1,\dots,x_n)} \,\leq\, a_n \\
    \frac{1}{|W|} \int_{W^n} M_1\big(\cap_{i=1}^n R(x_i)\big) \rho_n(x_1,\dots,x_n) \;\d{(x_1,\dots,x_n)} \,\leq\, b_n.
\end{align*}
In the case $d=2$, we additionally obtain
\begin{align*}
    &\frac{1}{|W|} \int_{W^n} \big|M_2\big(\cap_{i=1}^n R(x_i)\big)\big| \rho_n(x_1,\dots,x_n) \;\d{(x_1,\dots,x_n)} \,\leq\, c_n.
\end{align*}
Furthermore we have
\begin{align*}
    \sum_{n=1}^\infty \frac{a_n}{n!} \,<\, \infty, \qquad \sum_{n=1}^\infty \frac{b_n}{n!} \,<\, \infty, \qquad \sum_{n=1}^\infty \frac{c_n}{n!} \,<\, \infty.
\end{align*}
\end{Lem}

\begin{proof}
The statements are obvious for $n=1$, so we assume $n\geq 2$ in the following.
Let $j\in\{0,1\}$ or $j=2$ and $d=2$.
Then
\begin{align*}
    &\frac{1}{|W|} \int_{W^n} \big|M_j\big(\cap_{i=1}^n R(x_i)\big)\big| \rho_n(x_1,\dots,x_n) \;\d{(x_1,\dots,x_n)} \\
    &\qquad\qquad\qquad\qquad\qquad \,\leq\, \frac{1}{|W|} \int_W \int_{(\R^d)^{n-1}} \big|M_j\big(\cap_{i=1}^n R(x_i)\big)\big| \rho_n(x_1,\dots,x_n) \;\d{(x_2,\dots,x_n)} \;\d{x_1}.
\end{align*}
The function in the integral on the right side is translation invariant and therefore the inner integral is independent of $x_1$.
Replacing $x_1$ by $\0$ we obtain the upper bound
\begin{align} \label{nintegral}
    \int_{(\R^d)^{n-1}} \big|M_j\big(R(\0)\cap\cap_{i=2}^n R(x_i)\big)\big| \rho_n(\0,x_2,\dots,x_n) \;\d{(x_2,\dots,x_n)}.
\end{align}
The intersection $R(\0)\cap\cap_{i=2}^n R(x_i)$ is empty whenever one of the $x_i$ has distance greater than $2r$ from the origin.
Hence the integrand in \eqref{nintegral} vanishes outside $B(\0,2r)^{n-1}$.
For $j\in\{0,1\}$ we obtain by applying \eqref{VolumeSurface:upperBound} together with \eqref{Korfkt:uniformlybounded} the upper bound
\begin{align*}
    \int_{B(\0,2r)^{n-1}} nM_j(R(\0)) C^n \;\d{(x_2,\dots,x_n)} \,=\, n  C^n D^{n-1}M_j(R(\0)).
\end{align*}
In the case $j=2$ and $d=2$ we use \eqref{Euler:upperBound} and \eqref{Korfkt:uniformlybounded} to bound the integral from \eqref{nintegral} by
\begin{align*}
    \int_{B(\0,2r)^{n-1}} \left( \frac{n}{\pi s} M_{1}(R(\0)) + 8\binom{n}{2} \right) C^n \;\d{(x_2,\dots,x_n)} \,=\, c_n.
\end{align*}
Moreover, we have
\begin{align*}
    \sum_{n=1}^\infty \frac{a_n}{n!} \,=\, C\,M_0(R(\0))\sum_{n=1}^\infty \frac{(CD)^{n-1}}{(n-1)!} \,=\, C\,M_0(R(\0))\,\exp(CD).
\end{align*}
The same result holds upon replacing $a_n$ with $b_n$ and $M_0$ with $M_1$.
Finally, we get
\begin{align*}
    \sum_{n=1}^\infty \frac{c_n}{n!} \,&=\, \left( \frac{1}{\pi s}CM_1(R(\0)) + 4C^2D \right) \sum_{n=0}^\infty \frac{(CD)^n}{n!} \,=\, \left( \frac{1}{\pi s}M_1(R(\0)) + 4CD \right) C \exp(CD).
\end{align*}
\end{proof}

We now use Lemma~\ref{Lem:FUBDOM} to justify the applications of Fubini's theorem and the dominated convergence theorem in the derivation of the MCF representation.
In the following, we consider either $j\in\{0,1\}$ or $d=j=2$.
Let $\Phi$ be a stationary point process on $\R^d$ satisfying \eqref{Korfkt:uniformlybounded}, and let $\Phi_W$ denote its restriction to points in a measurable set $W\subset\R^d$.
According to Campbell's formula (Proposition 2.7 in \cite{LastPenrose16}), for any bounded and measurable set $W\subset\R^d$ with $|W|>0$, it holds that
\begin{align*}
    \E{ \sum_{x_1,\dots,x_n\in \Phi_W}^{\neq} M_j\left( \cap_{i=1}^n R(x_i) \right) } \,=\, \int_{W^n} M_j\left( \cap_{i=1}^n R(x_i) \right) \rho_n(x_1,\dots,x_n) \;\d{(x_1,\dots,x_n)},
\end{align*}
where $\sum^{\neq}$ indicates that the sum is taken over pairwise different points of $\Phi$.
By the inclusion--exclusion principle and Fubini's theorem, we obtain
\begin{align}
    \E{M_j\big( \cup_{x \in \Phi_W} R(x) \big)} \,&=\, \E{ \sum_{n=1}^\infty \frac{(-1)^{n+1}}{n!} \sum_{x_1,\dots,x_n\in \Phi_W}^{\neq} M_j\left( \cap_{i=1}^n R(x_i) \right) } \nonumber \\
    &=\, \sum_{n=1}^\infty \frac{(-1)^{n+1}}{n!} \int_{W^n} M_j\left( \cap_{i=1}^n R(x_i) \right) \rho_n(x_1,\dots,x_n) \;\d{(x_1,\dots,x_n)}, \label{eq:MCFinW}
\end{align}
where Lemma \ref{Lem:FUBDOM} verifies the conditions for the application of Fubini's theorem in the last equation.
In order to take the thermodynamic limit, we consider a sequence $(W_m)_{m \in \mathbb{N}}$ of convex sets such that the inradius $r(W_m)$ tends to infinity for $m\rightarrow\infty$.  
The inradius of a convex set $A$ is defined as the radius of the largest ball contained in $A$.
Due to Lemma \ref{Lem:FUBDOM} we can apply the dominated convergence theorem and use \eqref{eq:MCFinW} to obtain
\begin{align}
    \lim_{m\rightarrow\infty} \frac{1}{|W_m|} \E{M_j\big( \cup_{x \in \Phi_{W_m}} R(x) \big)} \,&=\, \sum_{n=1}^\infty \frac{(-1)^{n+1}}{n!} \lim_{m\rightarrow\infty} \frac{1}{|W_m|} \int_{W_m^n} M_j\left( \cap_{i=1}^n R(x_i) \right) \rho_n(x_1,\dots,x_n) \;\d{(x_1,\dots,x_n)} \nonumber \\
    &\hspace{-3cm}=\, \sum_{n=1}^\infty \frac{(-1)^{n+1}}{n!} \int_{(\R^d)^{n-1}} M_j\left( R(\0)\cap\cap_{i=2}^{n} R(x_i) \right) \rho_n(\0,x_2,\dots,x_{n}) \;\d{(x_2,\dots,x_{n})}. \label{eq:Formula_MCF_appendix}
\end{align}
For each fixed $n$, the second equality follows from Lemma~7.1 in \cite{Pabst.Euler}.
For $n=1$, the integral is to be interpreted as $M_j(R(\0))\rho_1(\0)$.
The integrals appearing in the resulting series are finite by Lemma~\ref{Lem:FUBDOM}.

\section{Proof of Theorem~\ref{Th:positiveReachIntersections}}\label{App:proof}

For $a,x\in\R^d$ with $\Vert x-a\Vert\in\{s,r\}$, that is $a\in\partial R(x)$, denote by $N(x,a)$ the unit normal vector of $R(x)$ at $a$.
Otherwise let $N(x,a):=\0$ (the origin in $\R^d$).
For $k\in\N$ define the function $f_k: (\mathbb{S}^{d-1})^k\rightarrow\R$ by
\begin{align*}
    f_k(u_1,\dots,u_k) \,:=\, \inf_{(t_1,\dots,t_k)\in\R^k_{\geq 0}\setminus\{\0\}} \frac{\Vert t_1u_1+\dots +t_ku_k\Vert}{t_1+\dots +t_k}.
\end{align*}
Here $\mathbb{S}^{d-1}$ denotes the unit sphere in $\R^d$.
The following lemma establishes some initial important properties of the functions $f_k$.

\begin{Lem} \label{Lem:Properties_fk}
\begin{enumerate}
    \item Using the notation $\Delta:=\{t\in [0,1]^k : \sum_{i=1}^k t_i = 1\}$ the function $f_k$ satisfies
    \begin{align*}
        f_k(u_1,\dots,u_k)=\inf_{(t_1,\dots,t_k)\in\Delta} \Vert t_1u_1+\dots +t_ku_k\Vert.
    \end{align*}
    \item The function $f_k$ is continuous.
    \item We have
    \begin{align} \label{Characterisation:positiveInf}
        f_k(u_1,\dots,u_k)>0 \quad\Longleftrightarrow\quad \exists w\in\mathbb{S}^{d-1}: \enskip \SKP w {u_i} > 0 \enskip\forall i\in \{1,\dots,k\}.
    \end{align}
\end{enumerate}
\end{Lem}

\begin{proof}
\begin{enumerate}
    \item The desired assertion follows from the fact for each $(u_1,\dots,u_k)\in(\mathbb{S}^{d-1})^k$ the function
    \begin{align*}
        g: \R^k_{\geq 0}\setminus\{\0\}\rightarrow\R, \quad g(t_1,\dots,t_k) \,:=\, \frac{\Vert t_1u_1+\dots +t_ku_k\Vert}{t_1+\dots +t_k}
    \end{align*}
    satisfies $g(\lambda t)=g(t)$ for all $\lambda>0$ and $t\in\R^k_{\geq 0}\setminus\{\0\}$.
    \item Using part (i) the Berge Maximum Theorem (Theorem 17.31 in \cite{Aliprantis}) shows that $f_k$ is continuous by choosing a constant correspondence $\varphi:(\mathbb{S}^{d-1})^k\rightarrow\mathcal{B}(\R^k)$, $\varphi(u)=\Delta$.
    \item It immediately follows from part (i) that
    \begin{align*}
        f_k(u_1,\dots,u_k) > 0 \quad\Longleftrightarrow\quad \0\notin \mathrm{conv}(u_1,\dots,u_k)=: S,
    \end{align*}
    where $\mathrm{conv}(u_1,\dots,u_k)$ denotes the convex hull of $\{u_1,\dots,u_k\}$.
    We conclude the proof by showing
    \begin{align*}
        \0\notin S \quad\Longleftrightarrow\quad \exists w\in\mathbb{S}^{d-1}: \enskip \SKP w {u_i}> 0 \enskip\forall i\in \{1,\dots,k\}.
    \end{align*}
    \begin{enumerate}
        \item[``$\Longrightarrow$'':] The Separation Theorem in \cite{Hug.ConvexGeometry} (Theorem 1.17) shows with the choice of $A=\{\0\}$ and $B=S$ that there is a $w\in\mathbb{S}^{d-1}$ such that $\SKP w x >0$ for all $x\in S$, which implies the desired result.
        \item[``$\Longleftarrow$'':] Let $x\in S$ and define $\varepsilon:=\min_{i\in\{1,\dots,k\}} \SKP w {u_i}>0$.
        Then $x$ can be written as $\sum_{i=1}^k t_iu_i$ with $t\in\Delta$ and we obtain
        \begin{align*}
            \SKP w x \,=\, \sum_{i=1}^k t_i \SKP w {u_i} \,\geq\, \varepsilon.
        \end{align*}
        Since this is true for all $x\in S$, it follows that $\0\notin S$.
     \end{enumerate}
\end{enumerate}
\end{proof}

Observe that the condition on the right-hand side of \eqref{Characterisation:positiveInf} means that $u_1, \dots, u_k$ lie in a single open half-space.
This is the case, for example, whenever $u_1, \dots, u_k$ are linearly independent, since in that case $\0 \notin \mathrm{conv}(u_1, \dots, u_k)$.
The following lemma shows that this linear independence holds for almost all configurations of ring centers.

\begin{Lem}\label{Lem:GeneralPosition}
For almost all $(x_1,\dots,x_n)\in(\R^d)^n$ the following holds: for every $a\in\R^d$ the vectors $a-x_i$, $i\in I(a):=\{i:a\in\partial R(x_i)\}$, are linearly independent.
In particular, no point lies on the boundaries of more than $d$ rings.
\end{Lem}

\begin{proof}
Fix a nonempty set $J\subseteq\{1,\dots,n\}$ and radii $t_i\in\{s,r\}$ for $i\in J$.
Consider the smooth map
\begin{align*}
F:\R^d\times(\R^d)^J&\,\longrightarrow\, \R^J,\\
(a,(x_i)_{i\in J})&\,\longmapsto\, \big(\Vert a-x_i\Vert^2-t_i^2\big)_{i\in J}.
\end{align*}
At each zero of $F$, the derivative of the $i$-th component with respect to $x_i$ is $v\mapsto -2\SKP{a-x_i}{v}$, which does not vanish since $\Vert a-x_i\Vert=t_i>0$, and its derivative with respect to $x_j$, $j\neq i$, is zero.
Hence $0$ is a regular value of $F$.
By the parametric transversality theorem (see~\cite[p.~68]{GuilleminPollack}), $0$ is a regular value of $F(\cdot,(x_i)_{i\in J})$ for almost all $(x_i)_{i\in J}$, that is, the gradients $2(a-x_i)$, $i\in J$, are linearly independent at every $a$ with $\Vert a-x_i\Vert=t_i$ for all $i\in J$.
Since there are only finitely many choices of $J$ and $(t_i)_{i\in J}$, the assertion follows.
\end{proof}

Using Lemmas~\ref{Lem:Properties_fk} and~\ref{Lem:GeneralPosition},
we can now prove Theorem~\ref{Th:positiveReachIntersections}.

\begin{proof}
    The key result for the proof is Theorem 4.19 in \cite{Rataj}.
    First we show that for almost all $x_1,\dots,x_n$ it is true that for each boundary point $a\in\partial (\cap_{i=1}^n R(x_i))$ we have
    \begin{align*}
        \eta(a) \,:=\, \inf\left\{ \frac{\Vert v_1+\dots+v_n\Vert}{\Vert v_1\Vert+\dots+\Vert v_n\Vert} : v_i\in\mathrm{Nor}(R(x_i),a), \Vert v_1\Vert+\dots+\Vert v_n\Vert>0 \right\} > 0.
    \end{align*}   
    We fix a boundary point $a\in\partial (\cap_{i=1}^n R(x_i))$ and define $I(a):=\{i\in\{1,\dots,n\} : a\in\partial R(x_i) \}$.
    Observe that $I(a)\neq\emptyset$.
    Without loss of generality, let $I(a)=\{1,\dots,k\}$.
    Since $\mathrm{Nor}(R(x_i),a)$ for $i\in I(a)$ is of the form $\{ tu_i : t\geq 0\}$ with $u_i=N(x_i,a)$, we have
    \begin{align*}
        \eta(a) \,=\, f_k(u_1,\dots,u_k)
    \end{align*}
    By Lemma~\ref{Lem:GeneralPosition}, for almost all $x_1,\dots,x_n\in\R^d$, the unit normal vectors $u_i:=N(x_i,a)$, $i\in I(a)$, are linearly independent at every boundary point $a\in\partial(\cap_{i=1}^n R(x_i))$.
    Consequently, they satisfy the right-hand side of \eqref{Characterisation:positiveInf}, and Lemma~\ref{Lem:Properties_fk} (iii) yields $\eta(a)>0$ at every such boundary point.
    
    Now fix $x_1,\dots,x_n\in\R^d$ such that $\eta(a)>0$ for all boundary points $a\in\partial (\cap_{i=1}^n R(x_i))$.
    It remains to show that $\eta$ is bounded away from zero uniformly over the boundary.
    We decompose the boundary $\partial (\cap_{i=1}^n R(x_i))$ into the sets
    \begin{align*}
        A_J \,:=\, \big\{ a\in\partial (\cap_{i=1}^n R(x_i)) : I(a)=J \big\}, \qquad \emptyset\neq J\subseteq\{1,\dots,n\}.
    \end{align*}
    Since $\partial R(x_i)$ is closed for every $i$, we have $J\subseteq I(a)$ for every $a\in\mathrm{cl}(A_J)$.
    For $J=\{i_1,\dots,i_k\}$ we consider the function
    \begin{align*}
        g_J: \mathrm{cl}(A_J) \rightarrow\R, \quad a\mapsto f_k\big( N(x_{i_1},a),\dots,N(x_{i_k},a)\big)
    \end{align*}
    on the closure $\mathrm{cl}(A_J)$ of $A_J$.
    The assumption implies $g_{I(a)}(a)>0$ for all $a\in\partial (\cap_{i=1}^n R(x_i))$, which also implies $g_J(a)>0$ for $J\subseteq I(a)$.
    Observe that $\mathrm{cl}(A_J)$ may contain points $a$ with $J\subseteq I(a)$, but $J\neq I(a)$.
    Together with Lemma \ref{Lem:Properties_fk} (ii) it follows that $g_J$ is a strictly positive and continuous function on a compact set.
    For every $J$ with $A_J\neq\emptyset$, the extreme value theorem therefore implies $\inf_{a\in\mathrm{cl}(A_J)}g_J(a)>0$.
    Consequently, we obtain
    \begin{align*}
        \inf \big\{ \eta(a) : a\in\partial (\cap_{i=1}^n R(x_i))\big\} \,=\, \inf_{\emptyset\neq J\subseteq \{1,\dots,n\}} \inf_{a\in A_J} g_J(a) \,>\,0.
    \end{align*}
    From Theorem 4.19 in \cite{Rataj}, we deduce that $\cap_{i=1}^n R(x_i)$ has positive reach.
\end{proof}

\section{Proof of Miles' formulas}
\label{App:MilesProof}

We conclude the appendix with the proof of Lemma~\ref{Lem:Miles}.
Let $\Phi$ be a stationary Poisson process on $\R^d$ with intensity $\gamma$.
Its $n$-th correlation function is given by
\begin{align*}
    \rho_n(x_1,\dots,x_n)=\gamma^n.
\end{align*}
We can therefore apply the general representation~\eqref{eq:Formula_MCF} and evaluate the resulting integrals explicitly for $M_0$ and $M_1$.

\begin{proof}
For the volume $M_0$ we obtain by applying Fubini's theorem
\begin{align*}
    \int_{(\R^d)^{n-1}} &M_0\left( R(\0)\cap R(x_2)\cap\dots\cap R(x_n) \right) \gamma^n \;\d{(x_2,\dots,x_{n})} \\
    &=\, \gamma^n \int_{(\R^d)^{n-1}} \int_{\R^d} \1\{y\in R(\0)\}\1\{y\in R(x_2)\}\cdot\,\dots\,\cdot\1\{y\in R(x_n)\} \;\d y \;\d{(x_2,\dots,x_{n})} \\
    &=\, \gamma^n \int_{\R^d} \1\{y\in R(\0)\} \int_{(\R^d)^{n-1}} \1\{x_2\in R(y)\}\cdot\,\dots\,\cdot\1\{x_n\in R(y)\} \;\d{(x_2,\dots,x_{n})} \;\d y \\
    &=\, \gamma^n M_0(R(\0))^{n-1} \int_{\R^d} \1\{y\in R(\0)\} \;\d y \,=\, \gamma^nM_0(R(\0))^n.
\end{align*}
The representation \eqref{eq:Formula_MCF} of $M_0(r,s,\Phi)$ now yields
\begin{align*}
    M_0(r,s,\Phi) \,=\, \sum_{n=1}^\infty \frac{(-1)^{n+1}}{n!} \gamma^nM_0(R(\0))^n \,=\, 1-\exp\left( -\gamma M_0(R(\0)) \right).
\end{align*}

Up to sets of $(d-1)$-dimensional Hausdorff measure zero, the boundary of an intersection of rings can be decomposed into subsets of the boundaries of the individual rings.
Denoting by $\alpha_i$ respectively $\alpha$ the $(d-1)$-dimensional Hausdorff measure on $\R^d$ restricted to the boundary of $R(x_i)$ respectively $R(\0)$, it is true that
\begin{align*}
    M_{1}\left( R(x_1)\cap\dots\cap R(x_n) \right) \,=\, \frac{1}{2} \sum_{i=1}^n \alpha_i\left( R(x_1)\cap R(x_2)\cap\dots\cap R(x_n) \right) \,=\, \frac{1}{2} \sum_{i=1}^n \alpha\big(R(x_1-x_i)\cap\dots\cap R(x_n-x_i)\big)
\end{align*}
and therefore with $x_1:=\0$
\begin{align*}
    \int_{(\R^d)^{n-1}} &M_{1}\left( R(\0)\cap R(x_2)\cap\dots\cap R(x_n) \right) \gamma^n \;\d{(x_2,\dots,x_{n})} \\
    &=\, \frac{1}{2}\gamma^n \sum_{i=1}^n \int_{(\R^d)^{n-1}} \alpha\big(R(-x_i)\cap R(x_2-x_i)\cap\dots\cap R(x_n-x_i)\big) \;\d{(x_2,\dots,x_{n})}.
\end{align*}
We provide an explicit computation of the summand for $i=1$.
The remaining summands can be computed in a similar way and yield the same result, although their expressions are slightly more cumbersome.
We again use Fubini's theorem and obtain
\begin{align*}
    \int_{(\R^d)^{n-1}} &\alpha\big(R(\0)\cap R(x_2)\cap\dots\cap R(x_n)\big) \;\d{(x_2,\dots,x_{n})} \\
    &=\, \int_{(\R^d)^{n-1}} \int_{\R^d} \1\{y\in R(\0)\} \1\{y\in R(x_2)\}\cdot\,\dots\,\cdot\1\{y\in R(x_n)\} \; \alpha(\d y) \;\d{(x_2,\dots,x_{n})} \\
    &=\, \int_{\R^d} \1\{y\in R(\0)\} \int_{(\R^d)^{n-1}} \1\{x_2\in R(y)\}\cdot\,\dots\,\cdot\1\{x_n\in R(y)\} \;\d{(x_2,\dots,x_{n})} \; \alpha(\d y) \\
    &=\, M_0(R(\0))^{n-1} \alpha(R(\0)) \,=\, 2M_0(R(\0))^{n-1}M_{1}(R(\0)).
\end{align*}
Altogether, we obtain
\begin{align*}
    \int_{(\R^d)^{n-1}} M_{1}\left( R(\0)\cap R(x_2)\cap\dots\cap R(x_n) \right) \gamma^n \;\d{(x_2,\dots,x_{n})} \,=\, \gamma^n n M_0(R(\0))^{n-1}M_{1}(R(\0))
\end{align*}
and therefore by \eqref{eq:Formula_MCF}
\begin{align*}
    M_{1}(r,s,\Phi) \,=\, \gamma M_{1}(R(\0)) \sum_{n=1}^\infty \frac{(-1)^{n-1}}{(n-1)!} \gamma^{n-1} M_0(R(\0))^{n-1} \,=\, \gamma M_{1}(R(\0)) \exp\left( -\gamma M_0(R(\0)) \right).
\end{align*}
\end{proof}

\bibliographystyle{apsrev4-2}
\bibliography{LiteratureMCF}

\end{document}